\documentclass[11pt,a4paper]{article}
\pdfoutput=1

\usepackage{mymacro-jhep}

\renewcommand{\jb}{{\bar{\jmath}\llsp}}
\newcommand{\un}[1]{\underline{#1}}
\renewcommand{\geq}{\geqslant}

\newcommand{\DQp}{D_Q^+}
\newcommand{\DQm}{D_Q^-}
\newcommand{\DQ}[1]{D_Q^{#1}}
\newcommand{\DQbp}{D_\Qb^+}
\newcommand{\DQbm}{D_\Qb^-}
\newcommand{\DQb}[1]{D_\Qb^{#1}}
\newcommand{\DQsq}{D_{Q^2}}
\newcommand{\DQbsq}{D_{\Qb^2}}
\newcommand{\DQQbpp}{D_{Q\Qb}^{++}}

\newcommand{\DQQb}[2]{D_{Q\Qb}^{#1#2}}

\newcommand{\DQsqQb}[1]{D_{Q^2\Qb}^{#1}}

\newcommand{\DQbsqQ}[1]{D_{\Qb{}^2Q}^{#1}}
\newcommand{\DQsqQbsq}{D_{Q^2\Qb{}^2}}
\newcommand{\DQlQblb}{D_{Q^\ell\Qb{}^\ellb}}
\newcommand{\CDQp}{\CD_Q^+}
\newcommand{\CDQm}{\CD_Q^-}
\newcommand{\CDQ}[1]{\CD_Q^{#1}}
\newcommand{\CDQbp}{\CD_\Qb^+}
\newcommand{\CDQbm}{\CD_\Qb^-}
\newcommand{\CDQb}[1]{\CD_\Qb^{#1}}

\newcommand{\CDQQb}[2]{\CD_{Q\Qb}^{#1#2}}
\newcommand{\CDQlbQbl}{\CD_{Q^\ellb\Qb{}^\ell}}

\newcommand{\CXb}{\overbar{\mathcal{X}}}

\newcommand{\mathematicapack}{https://gitlab.com/maneandrea/spinoralgebra/tree/master}

\author{Andrea Manenti}

\affiliation{
Institute of Physics, École Polytechnique Fédérale de Lausanne, CH-1015 Lausanne, Switzerland}
\affiliation{
Simons Center for Geometry and Physics, Stony Brook, NY 11794, USA}

\emailAdd{andrea.manenti@epfl.ch}

\title{Differential operators for superconformal correlation functions}

\abstract{We present a systematic method to expand in components four dimensional superconformal multiplets. The results cover all possible $\mathcal{N} = 1$ multiplets and some cases of interest for $\mathcal{N} = 2$. 
As an application of the formalism we prove that certain $\mathcal{N} = 2$ spinning chiral operators (also known as ``exotic'' chiral primaries) do not admit a consistent three-point function with the stress tensor and therefore cannot be present in any local SCFT. This extends a previous proof in the literature which only applies to certain classes of theories. To each superdescendant we associate a superconformally covariant differential operator, which can then be applied to any correlator in superspace. In the case of three-point functions, we introduce a convenient representation of the differential operators that considerably simplifies their action. As a consequence it is possible to efficiently obtain the linear relations between the OPE coefficients of the operators in the same superconformal multiplet and in turn streamline the computation of superconformal blocks. We also introduce a Mathematica package to work with four dimensional superspace.}

\begin{document}

\maketitle

\section{Introduction}

Superconformal field theories (SCFTs) have a remarkably rigid structure. The superconformal algebras have been classified long ago~\cite{Nahm:1977tg} and their unitary representations have been studied extensively, leading to a classification in four dimensions~\cite{Dobrev:1985qv, Dolan:2002zh} and in any dimension~\cite{Cordova:2016emh, Buican:2016hpb}. A mathematical obstruction prevents superconformal algebras to exist in dimensions greater than six; furthermore, assuming that the resulting SCFT is local puts an upper bound on the amount of supercharges $\CN$.\footnote{In three dimensions there actually exist theories for any $\CN$, but those for $\CN > 8$ are necessarily free.} These facts made the classification of unitary representations an attainable task as only a finite number of cases needed to be considered. At times, however, studying superconformal multiplets purely from a group theoretical point of view might not be enough. Often one needs to know how to relate in a precise way the correlation functions of the various operators in the same multiplet in order to fully exploit the constraints of supersymmetry. This can be done, for example, with the help of a superspace formulation. In the case of four dimensions, which is the one we will focus on, such a formalism has been introduced in the end of the nineties~\cite{Park:1997bq,Osborn:1998qu,Park:1999pd}.
\par
A very general and successful method for studying superconformal field theories is the conformal bootstrap~\cite{Rattazzi:2008pe, Poland:2018epd}.\footnote{By conformal bootstrap we mean the modern revival of the original bootstrap program, which was developed in the early seventies \cite{Ferrara:1973yt,Polyakov:1974gs,PhysRevD.13.887,Dobrev:1977qv}.} It was initially developed without having supersymmetry in mind, but its application to SCFTs followed soon after~\cite{Poland:2010wg, Vichi:2011ux}. The search for four dimensional theories with $\CN=1$ supersymmetry has been mainly focused on the recently discovered ``minimal'' SCFT~\cite{Poland:2011ey, Poland:2015mta, Li:2017ddj}. But the bootstrap is certainly not limited to four supercharges. There are examples in $\CN=2$~\cite{Beem:2014zpa, Lemos:2015awa}, including a study of the $(A_1,A_2)$ Argyres-Douglas models~\cite{Cornagliotto:2017snu}, and there are also works addressing $\CN=4$ super Yang-Mills~\cite{Beem:2013qxa, Beem:2016wfs}. We should mention that this program has been very successful in other spacetime dimensions as well. The studies on $(2,0)$ theory in six dimensions~\cite{Beem:2015aoa} and the recent works on three-dimensional SCFTs relevant for $M$-theory are some notable examples~\cite{Agmon:2017xes, Agmon:2019imm}.
\par
All these works impose the bootstrap constraints on a multiplet whose superprimary is a Lorentz scalar. Most of the results involving supersymmetric conformal blocks indeed focus on scalar external operators~\cite{Nirschl:2004pa, Fortin:2011nq, Fitzpatrick:2014oza, Khandker:2014mpa, Li:2016chh, Ramirez:2016lyk, Li:2018mdl, Ramirez:2018lpd}. In the recent years, however, a lot of progress has been made towards developing a powerful formalism for spinning operators. The process of computing conformal blocks is fully automated in $3\llsp d$~\cite{Erramilli:2019njx} and the simplest case of spin $1/2$ has been already tackled in $4\llsp d$~\cite{Karateev:2019pvw}. For all other Lorentz representations at least two approaches are possible: the method of weight-shifting differential operators~\cite{Echeverri:2016dun, Karateev:2017jgd} and the recently developed formalism in embedding space~\cite{Fortin:2019fvx, Fortin:2019dnq, Fortin:2019gck}, for which there are already some explicit results expressed in terms of specific substitutions on Gegenbauer polynomials.
Soon it will be possible to address supersymmetric theories with a bootstrap setup that involves spinning superconformal primaries. The necessary ingredients are already available for the Ferrara-Zumino multiplet in $\CN=1$~\cite{Manenti:2018xns}. On the other hand, for other Lorentz representations in $\CN=1$ or for non-scalar primaries in $\CN \geq 2$ there are no explicit results so far. Recently a general theory has been developed~\cite{Buric:2019rms}. In the formalism of~\cite{Buric:2019rms} the superconformal blocks are computed by solving the Casimir equation. The problem can be translated into the Schr\"odinger equation of a Calogero-Sutherland model~\cite{Isachenkov:2016gim}, or rather, a perturbation of it which becomes exact at a finite order. In this paper, however, we propose a different approach to the computation of $\CN=1,2$ superconformal blocks. The final output of the formalism shown here will be a set of linear relations among the OPE coefficients and norms of the operators in the same multiplet\footnote{We call $\lambda$ the coefficient appearing in the three-point function in some standard basis and $\CC$ the normalization of the two-point function relative to some standard convention. See \sectionname~\ref{sec:packconvention} for more details.}
\eqn{
\lambda^{(a)}_{(Q^\ell\Qb{}^{\ellb}\CO_1)\CO_2\CO_3} = M^a_{\phantom{a}b}\, \lambda^{(b)}_{\CO_1\lnsp\CO_2\CO_3}\,,\qquad
\CC_{(Q^\ell\Qb{}^{\ellb}\CO_1)} = N \,\CC_{\CO_1}\,,
}
for some, in general rectangular, matrix $M$ and some complex number $N$. Once the conformal blocks of all operators in a multiplet are known, it suffices to take the appropriate linear combination following from the knowledge of $M$ and $N$.
\par
Even though the main motivation behind this paper is that of computing superconformal blocks, there are other interesting applications that can be studied. We first directed our attention to the so-called ``exotic chiral primaries.'' These operators are allowed by representation theory\footnote{The exotic operators have spin $(j,0)$ with $j>0$ and are annihilated by the $\Qb$ supercharges. In the notation of~\cite{Dolan:2002zh} they are $\overbar{\CE}_{\frac{r}2(j,0)}$ and in the notation of \cite{Cordova:2016emh} they are $L\overbar{B}_1[j;0]_{r/2}^{0,r}$.} but have been proven to be absent from a very large class of theories~\cite{Buican:2014qla}. Arguments that excluded them from theories of class $\CS$ appeared in~\cite{Gadde:2011uv}, while the more general result of~\cite{Buican:2014qla}  applies to $i)$ theories with a Lagrangian description, $ii)$ theories related to Lagrangian ones via a generalized Argyres-Seiberg-Gaiotto duality and $iii)$ theories that flow to an IR Lagrangian theory via an $\CN=2$--preserving deformation. We further extend these results showing that the exotic chiral primaries are forbidden in any local SCFT. Using the formalism presented here\footnote{We should point out that the differential operator needed for the proof, while being a particular case of the ones defined in this paper, was already known from~\cite{Kuzenko:1999pi}.} we expand a three-point function of an exotic primary $\CX$, its conjugate $\CXb$ and the stress tensor $\CJ$ into $\CN=1$ multiplets. We then show that the resulting correlators do not satisfy the Ward identities computed in~\cite{Manenti:2019kbl} and are therefore inconsistent. The only assumption is the existence of a conserved stress tensor, i.e. locality.
\par
Another possible application is the investigation of the constraints stemming from the averaged null energy condition (ANEC). The ANEC~\cite{Hofman:2008ar, Faulkner:2016mzt, Hartman:2016lgu} states the positivity of integrated expectation values of the stress tensor. There are non-supersymmetric results only in four dimensions for spin $(j,0)$ and $(j,1)$~\cite{Cordova:2017dhq} and $\CN=1$ supersymmetric results for general multiplets with spin $(j,0)$~\cite{Manenti:2019kbl}, which impose lower bounds on the conformal dimension, sometimes stronger than unitarity. As of now, there is no intuition on the shape of these bounds for more general Lorentz representations or for extended supersymmetry and thus more investigations are needed. 
\par
The approach of this paper consists in defining a set of superconformally covariant differential operators that can be applied to any correlator in superspace. In $\CN=1$, by setting the Grassmann variables to zero, one obtains a conformal primary. Whereas in $\CN=2$, by setting to zero only the Grassmann variables $\theta_{\un2},\thetab^{\un2}$, one obtains an $\CN=1$ superconformal primary. The advantage of the operators that we define is that it is possible to greatly simplify their action on three-point functions. This is because any three-point function can be factorized as a prefactor and a function $t$ of one point in superspace $Z$. Then our differential operators can be mapped to derivatives on $t(Z)$, reducing the problem to a single point. In order to define such operators we need to carefully subtract the conformal descendants --- or the superdescendants of the other supercharges in the $\CN=2$ case --- that are generated when acting with $Q,\,\Qb$ on $\CO$. This is done in full generality to all orders in $\CN=1$. On the other hand, only some cases have been considered in $\CN=2$, leaving the complete analysis to a future work.
Specifically we considered all operators with vanishing $\mathfrak{su}(2)$ R-charge up to quadratic order in the $Q^{\un2},\,\Qb_{\un2}$ supercharges and all operators with $\mathfrak{su}(2)$ R-charge $1$ and $1/2$ up to linear order. By expanding the differentiated three-point function in a standard basis one can read out the linear relations among the OPE coefficients of $\CO$ and $Q^\ell\Qb{}^\ellb\CO$. Similarly, by acting on two-point functions, one can obtain the relative norms, even though they are already known in general for $\CN=1$~\cite{Li:2014gpa}. We also introduce a Mathematica package to work on four dimensional superspace. It can be used to derive the results presented in this paper and to perform the computation of superconformal blocks of spinning primaries. It can also be regarded as a general purpose tool for any tensor computation in four dimensions with the index-free formalism.
\par
Finally we should mention that superconformal differential operators for extended supersymmetry have been introduced in the past~\cite{Dobrev:1985qz}. The operators that we define in this paper are unrelated to those.
\par
The paper is organized as follows. In \sectionname~\ref{sec:review} we give a general overview of the superspace formalism both for $\CN=1$ and $\CN=2$, leaving the precise definitions and some details to \appendixname~\ref{app:notation}. Then we present all the differential operators in \sectionname~\ref{sec:diffop}. In \sectionname~\ref{sec:threepf} we focus on three-point functions and study the action of the differential operators on them. In \sectionname~\ref{sec:exotic} we present the proof of the absence of exotic primaries from any local SCFT. Finally, in \sectionname~\ref{sec:mathematica} we introduce the Mathematica package used for this paper, discuss its applications and show a small worked out example. Further details are left to the appendices.

\section{Review of the formalism}\label{sec:review}

\subsection{Notation and conventions}

We will follow the conventions of \cite{WessnBagger} for four dimensional spinors and utilize the formalism of \cite{Park:1997bq, Osborn:1998qu} for $\CN=1$ superspace and its generalization \cite{Park:1999pd, Kuzenko:1999pi} for $\CN=2$ superspace.\par
We contract all $\alpha,\alphad$ indices with commuting spinors $\eta^\alpha, \etab^\alphad$ and we use the shorthand $\bfx$ to denote the tuple $x,\eta,\etab$. An operator of spin $(j,\jb)$ is denoted as follows
\eqn{
O(\bfx) \equiv O(x,\eta,\etab) = \eta^{\alpha_1}\cdots \eta^{\alpha_j} \,\etab^{\alphad_1}\cdots \etab^{\alphad_\jb} \,O_{(\alpha_1\cdots\alpha_j)\,(\alphad_1\cdots\alphad_\jb)}(x)\,.
}[]
An operator in superspace also depends on Grassmann variables $\theta_I^\alpha, \thetab^I_\alphad$, where $I=1,2$ is an $\mathfrak{su}(2)_R$ index, only present in the $\CN=2$ case. We use a shorthand $z$ to denote the tuple $x,\theta_I,\thetab^I$, and $\bfz$ to denote $z,\eta,\etab$.
\eqn{
\CO(\bfz) \equiv \CO(x,\theta_I,\thetab^I,\eta,\etab) = e^{i \theta_I\llsp Q^I+ i \thetab^I \llsp \Qb_I}\,O(\bfx)\,.
}[defsuperspace]
Additional lower case indices $\eta_i,\theta_i,x_{ij}$, label the $i$-th operator in a correlation function. For more details on the notation and the definition of the supersymmetric intervals $\rmx_{i\jb}$, $\theta^\alpha_{I\,ij}$, $\thetab^I_{\alphad\,ij}$ see \appendixname\xspace \ref{app:notation}.
\par 
The various $\CN=1$ superdescendants are indicated as $(Q^\ell\Qb{}^{\bar\ell}O)$, for $\ell,\bar\ell=0,1,2$. By definition they are conformal primaries. We follow the conventions of~\cite{Li:2014gpa} for the normalization of their two-point function. When there are different choices for the spin, they are indicated with a $\pm$ superscript. E.g. $(Q\Qb O)^{+-}$ has spin $(j+1,\jb-1)$. In order to avoid confusion between exponents and possible explicit $\mathfrak{su}(2)_R$ indices, the latter will be underlined. Namely $(Q^{\underline{2}}O) \equiv (Q^{I=2}O)$ and $(Q^2O) \equiv (Q^\alpha Q_\alpha O)$. Following~\cite{Cordova:2016emh} we will define the $\mathfrak{u}(1)_R$ charge $r$ for both $\CN=1,2$ as
\eqn{
[r,Q^I_\alpha] = - Q^I_\alpha\,,\qquad [r,\Qb_{I\alphad}] = \Qb_{I\alphad}\,,
}[]
and the $\mathfrak{su}(2)_R$ charge $R$ as the Dynkin label $R \in \mathbb{N}$, with the Cartan $R_3$ taking values $-R/2,\ldots,R/2$. It will be convenient to define the $q,\qb$ charges as a function of $r$ and the scaling dimension $\Delta$ as follows
\eqna{
\Delta &= q+ \qb \,,&\qquad r &= \tfrac23(q-\qb)\,, \qquad &\mbox{for $\CN = 1$}\,,\\
\Delta &= q+ \qb \,,& r &= 2\llsp(q-\qb)\,, &\mbox{for $\CN = 2$}\,.\\
}[qqbcharges]
The unitarity bounds for long multiplets in these variables read
\eqna{
2\llsp q &\geq  j + 2 \,,&\qquad 2 \llsp \qb &\geq  \lsp \jb + 2\,, \qquad &\mbox{for $\CN = 1$}\,,\\
2\llsp  q &\geq j + 2 + R \,,&\qquad 2 \llsp \qb &\geq \jb + 2 + R\,, \qquad &\mbox{for $\CN = 2$}\,.\\
}[]
In the notation of~\cite{Cordova:2016emh} an $A$-type shortening happens when one of the inequalities is saturated and a $B$-type shortening happens when $j$ or $\jb$ vanishes and, respectively, $q$ or $\qb$ equals $R/2$ (or vanishes in the $\CN=1$ case). We will also choose to embed $\CN=1$ into $\CN=2$ by taking the subalgebra generated by $Q_\alpha^{\un1}$ and $\Qb_{\un1\alphad}$. This leads to
\eqn{
q_{\CN=1} = q_{\CN=2} - R_3\,,\qquad
\qb_{\CN=1} = \qb_{\CN=2} + R_3\,.
}[]
In the following we will define differential operators denoted with the letter ``$\!\llsp D\lsp$'' in several different fonts. In order not to generate confusion we summarize here their meaning. See the next subsections for the definition of the $Z$ variables.
\begin{table}[H]
\centering
\begin{tabular}{ccc}
Symbol & Supercharges & Acts on \\
\hline
$\DQlQblb$ & $\CN=1$ & $z_1 \, / \, z_2$ \\
$\CDQlbQbl \; / \;\CQ_{Q^\ellb\Qb{}^\ell}$ & $\CN=1$ & $Z_3$ \\
$\BBD_{Q^\ell\Qb{}^\ellb}$ & $\CN=2$&  $z_1 \, / \, z_2$ \\
$\mathfrak{D}_{Q^\ellb\Qb{}^\ell} \; / \; \mathfrak{Q}_{Q^\ellb\Qb{}^\ell}$ & $\CN=2$ & $Z_3$ \\
\hline
\end{tabular}
\label{tab}\caption{Reminder for the notation of differential operators. The alternatives in the first column represent the differential operator obtained by acting on the first / second operator, respectively.}
\end{table}

\subsection[\texorpdfstring{$\CN=1$}{N = 1} superspace]{$\boldsymbol{\CN=1}$ superspace}\label{sec:N1superspace}

Given three superconformal primaries $\CO_1, \CO_2$ and $\CO_3$ whose sum of R-charges is $0,1$ or $2$ in absolute value, one can define a three-point function as
\eqn{
\langle \CO_1(\bfz_1)\lsp\CO_2(\bfz_2)\lsp\CO_3(\bfz_3)\rangle = \CK_{\CO_1\CO_2}(\bfz_{1,2},z_3;\partial_{\chi_{1,2}},\partial_{\chib_{1,2}})\,t^{\CO_1\CO_2}_{\CO_3}(Z_3;\chi_{1,2},\chib_{1,2};\eta_3,\etab_3)\,.
}[GeneralThreepf]
The $\CK_{\CO_1\CO_2}$ is an universal prefactor and the $t^{\CO_1\CO_2}_{\CO_3}$ encodes all the information of the three-point function and can be expressed as a linear combination of tensor structures. The commuting spinors $\chi_i,\chib_i$ are auxiliary polarizations that are removed by the derivatives in the prefactor. The variable $Z_3$ collectively denotes the superconformally covariant variables $X_3,\Theta_3$ and $\Thetab_3$ whose definition can be found in \eqref{XThThbdef}. Clearly, since here the operator $\CO_3$ is treated differently, there are two other equivalent representation related by cyclic permutations. For more details see \appendixname~\ref{app:thirdpoint}.
\par
The general form of the prefactor $\CK_{\CO_1\CO_2}$ is the following
\eqn{
\CK_{\CO_1\CO_2} = \frac{1}{j_1\lnsp!\lsp\,\jb{}_1\lnsp!\lsp\,j_2!\lsp\,\jb{}_2!}\frac{\prod_{i=1}^2(\eta_i\lsp\rmx_{i\bar3}\lsp\partial_{\chib_i})^{j_i}\, (\partial_{\chi_i}\rmx_{3\ib}\lsp\etab_i)^{\jb_i}}{\prod_{i=1}^2 {{x_{\bar3i}}^{2q_i+j_i}\, x_{\ib3}}^{2\qb_i+\jb_i}}\,.
}[Kdef]
For the definition of $x_{i\jb}$ see \eqref{xijdef}. Note that, as the name suggests, the prefactor only depends on the quantum numbers of the first two operators.
The $t^{\CO_1\CO_2}_{\CO_3}$ can contain all Lorentz invariant combinations of its arguments as tensor structures. They need to be homogeneous functions of the auxiliary spinors --- with the degree dictated by the spins $j_i$ and $\jb_i$ --- and also satisfy a scaling property illustrated in \eqref{scalingt}. Due to the Schouten identities\footnote{The Schouten identities are all those that follow from $
\epsilon^{\alpha\beta} \epsilon^{\gamma \delta} + \epsilon^{\gamma\alpha} \epsilon^{\beta \delta} + \epsilon^{\beta\gamma} \epsilon^{\alpha \delta} = 0\,,
$
and the corresponding one with dotted indices.} these tensor structures can be hard to enumerate. But the expected number can be computed by a group theoretic formula~\cite{Manenti:2018xns}. Moreover, since the problem is essentially analogous to listing tensor structures in embedding space, one can easily obtain them by using the results of \cite{Elkhidir:2014woa,Cuomo:2017wme}. The idea is to first define the mapping
\eqna{
\hat{\mathbb{K}}^{ij}_k \,&\longrightarrow\, \eta_i\eta_j\,,
&\qquad\hat{\overline{\mathbb{K}}}^{ij}_k \,&\longrightarrow\, \etab_i\etab_j\,,\qquad  \\
\hat{\mathbb{J}}^i_{jk}\,&\longrightarrow\, U_3^{-1}\,\eta_i\mathrm{U}_3\lsp\etab_i\,
&\hat{\mathbb{I}}^{ij}\,&\longrightarrow\, U_3^{-1}\,\eta_i\mathrm{U}_3\lsp\etab_j\,,
}[mappingcfts4d]
where we have renamed $\chi_i$ and $\chib_i$ to $\eta_i$ and $\etab_i$ respectively for simplicity of notation and used $\mathrm{U}_3$ defined in \eqref{Udef}. Then we proceed order by order in $\Theta_3,\Thetab_3$. The order zero is trivial as it suffices to apply \eqref{mappingcfts4d} to the three-point function. Now say we want to compute the order $\Theta_3\Thetab_3$. We simply consider the three-point functions with all four combinations of spin\footnote{If $j_3$ or $\jb_3$ is zero the corresponding negative shift is omitted. The choice of shifting the spin labels of the third operator is unimportant and equivalent to any other choice, even when $j_3\jb{}_3 = 0$.}
\eqn{
(j_1,\jb_1),\, (j_2,\jb_2),\, (j_3 \pm 1, \jb_3\pm 1)\,,
}[]
and $\Delta_3 \to \Delta_3+1$. Then, after applying the mapping \eqref{mappingcfts4d}, we remove the extra $\eta_3,\etab_3$ spinors with $\Theta_3\partial_{\eta_3}$, $\Thetab_3\partial_{\etab_3}$ and attach missing $\eta_3,\etab_3$ spinors with $\Theta_3\eta_3$, $\Thetab_3\etab_3$. For quadratic orders it suffices to attach an overall $\Theta^2_3$ or $\Thetab{}^2_3$ to the three-point function with no shifts in the spins. The same logic applies to the other orders. The constraints of multiplet shortening and conservation can be applied directly on the $t$ by using the fact that the shortening differential operators always annihilate the prefactor with the appropriate quantum numbers. This is a consequence of
\eqna{
\mbox{$A_1$ shortening:}\qquad&\frac{\partial}{\partial\eta_{1\alpha}}D_{1\alpha} \frac{(\eta_1\rmx_{1\bar{3}}\etab_1)^j}{{x_{\bar{3}1}}{\!}^{2j+2}}\, f(x_{\bar13}) = 0\,,\\
\mbox{$A_2$ shortening:}\qquad&{D_1}{\!}^2 \frac{1}{{x_{\bar{1}3}}{\!}^2}\, f(x_{\bar13}) = 0\,,\\
\mbox{$B_1$ shortening:}\qquad& D_{1\alpha}\, f(x_{\bar13})=0\,,
}[shortening]
for $x_{13}\neq 0$. Similar identities hold for $\Db$. If $x_1=x_3$ we have the usual contact term singularity. Once the differential operator is past the prefactor we can use \eqref{DonF} to act on $t_{\CO_3}^{\CO_1\CO_2}$.

\subsection[\texorpdfstring{$\CN=2$}{N = 2} superspace]{$\boldsymbol{\CN=2}$ superspace}

With a formalism similar to the previous case one can construct superconformal three-point function of $\CN=2$ primaries. First we need to introduce the following unitary matrix
\eqn{
u_I^{\phantom{I}\lsp J}(z_{ij}) = \delta_I^{\phantom{I}\lsp J} - 4i\lsp \frac{\theta_{ij\lsp I}\lsp\rmx_{\ib j}\lsp\thetab_{ij}^J}{{x_{\ib j}}^2}\,.
}[uijdef]
By rescaling $u(z_{ij})$ appropriately we obtain a unimodular matrix
\eqn{
\hat{u}_I^{\phantom{I}\lsp J}(z_{ij}) = \left(\frac{{x_{\jb i}}^2}{{x_{\ib j}}^2}\right)^{\frac12}u_I^{\phantom{I}\lsp J}(z_{ij})\,,\qquad \hat{u}(z_{ij}) \in \mathrm{SU}(2)\,.
}[]
Let us consider three superconformal primaries $\CO_1^{\CI_1}$, $\CO_2^{\CI_2}$ and $\CO_3^{\CI_3}$. Here $\CI_i$ is an $\mathfrak{su}(2)$ index transforming under the representation $R_i$. Let us denote as
\eqn{
\CT_{\phantom{R}\;\,\CI}^{R\phantom{\CI\;}\CJ}(u)\,,\qquad
u\in\mathrm{SU}(2)\,,
}[]
the representation with Dynkin label $R$ of $\mathrm{SU}(2)$. The simplest cases are\footnote{$(\sigma^A)_{\phantom{J}I}^{J}$ are the usual three-dimensional Pauli matrices and $\epsilon_{\un1\un2} = \epsilon^{\un2\un1} = 1$ is the Levi-Civita tensor.}
\eqn{
\CT_{\phantom{1}\,I}^{\mathbf1\phantom{I}\lnsp J}(u) = u_I^{\;J}\,,
\qquad
\CT_{\phantom{1}\,A}^{\mathbf2\phantom{A}B}(u) = \frac14\lsp(\sigma^A \epsilon)^{J_1 I_1}(\epsilon\sigma^B)_{J_2 I_2}\big(u_{I_1}^{\;\,I_2}\,\lsp u_{J_1}^{\;\,J_2} + u_{I_1}^{\;\,J_2} \,\lsp u_{J_1}^{\;\,I_2}\big)\,.
}[ThalfTone]
The most general three-point function then has the following form
\eqna{
\langle \CO_1^{\CI_1}(\bfz_1)\lsp\CO_2^{\CI_2}(\bfz_2)\lsp\CO_3^{\CI_3}(\bfz_3)\rangle = \;&\CK_{\CO_1\CO_2}(\bfz_{1,2},z_3;\partial_{\chi_{1,2}},\partial_{\chib_{1,2}})\times\\&
\CT_{\phantom{R}\;\,\CI_1}^{R\phantom{\CI_1}\CJ_1}(\hat{u}(z_{13}))\,
\CT_{\phantom{R}\;\,\CI_2}^{R\phantom{\CI_2}\CJ_2}(\hat{u}(z_{23}))\times
\\&t^{\CO_1\CO_2}_{\CO_3}{}^{\left.\phantom{\int\hspace{-1.5ex}}\right|\,\CJ_1\CJ_2\llsp\CI_3}(Z_3;\chi_{1,2},\chib_{1,2};\eta_3,\etab_3)\,,
}[GeneralThreePFN2]
with $\CK$ defined in \eqref{Kdef} and $Z_3$ denoting $X_3,\Theta_3^I,\Thetab_{3\lsp I}$ (see \appendixname~\ref{app:notation}). The $t$ has the same scaling properties as the $\CN=1$ case (see \eqref{scalingt}). But in addition it has to transform as a tensor with the indices in the appropriate representations. The dependence on the $\mathfrak{su}(2)$ indices may come from $\Theta^I$, $\Thetab_I$ or explicit $\epsilon_{IJ}$ and $\delta^I_J$ tensors.

\section{Constructing the differential operators}\label{sec:diffop}

\subsection[\texorpdfstring{$\CN=1$}{N = 1} case]{$\boldsymbol{\CN=1}$ case}
\subsubsection{Strategy}

\begin{figure}[tbp]
\centering
\includegraphics[scale=1.2]{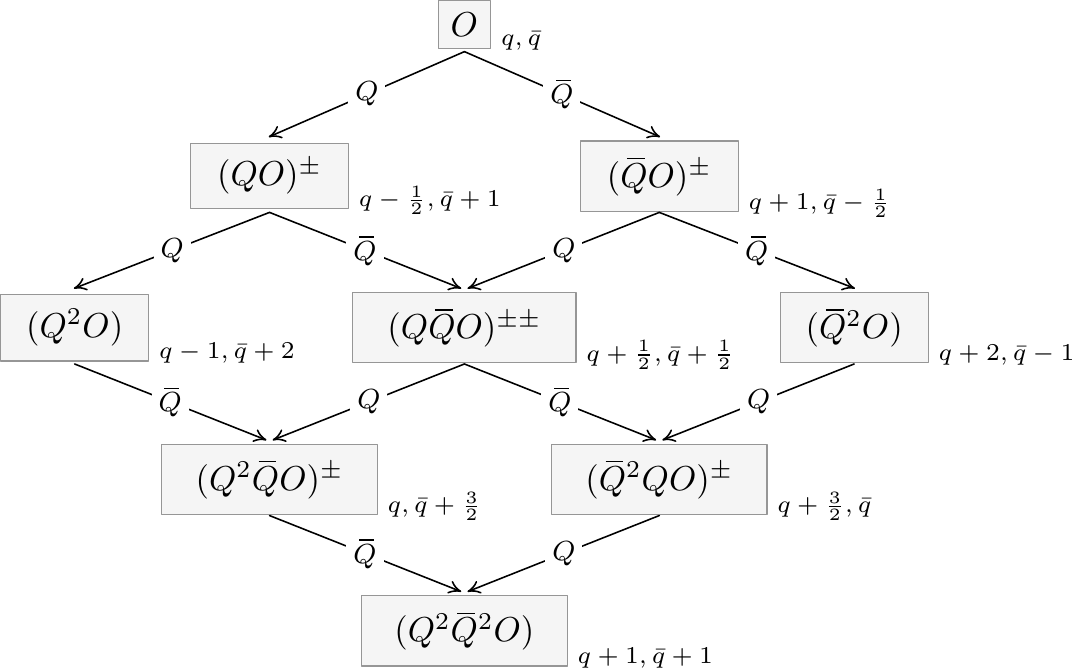}
\caption{Diagram of all operators in an $\CN=1$ long multiplet. Superscripts of $\pm$ indicate the choice of spin $j\pm1$ or $\jb\pm1$. Subscripts indicate the $q,\lsp\qb$ charges. The R-charge grows from left to right by $1$ and the conformal dimension grows from top to bottom by $1/2$. A box represents the full tower of descendants $O, \partial_\mu O, \partial^2 O, \ldots$}
\label{fig:N1}
\end{figure}
 
The goal of this section is to derive a set of superconformally covariant differential operators $D$ that extract a given order in $\theta,\thetab$ from an $\CN=1$ superconformal multiplet $\CO(\bfz)$. That is, we want them to satisfy the following property
\eqn{
\DQlQblb\,\CO(\bfz)\big|_0 = (Q^\ell\Qb{}^\ellb O)(\bfx)\,.
}[generalDllbdef]
where $|_0$ means evaluating at $\theta = \thetab = 0$ after taking the derivative. The set of all operators in a long multiplet is illustrated in \figurename~\ref{fig:N1}. 
Clearly, by the definition of $\CO(\bfz)$ \eqref{defsuperspace}, the first orders will be simply a derivative with respect to $\theta$ or $\thetab$. However, the situation becomes more complicated when both $\ell$ and $\ellb$ are nonzero. In this case the term multiplying $\theta^\ell\thetab{}^\ellb$ is not a conformal primary, but a linear combination of $(Q^\ell\Qb{}^{\ellb}O)$ and the descendants of the previous orders. Thus we need to be able to disentangle these contributions. Furthermore we will not be content with any form of the differential operator. We will need to express it as a combination of chiral and antichiral derivatives $D_\alpha$, $\Db_\alphad$ \eqref{chiralDDbdef}. The reason will be evident in the next section: these derivative have nice covariant properties that allow us to pass them through the prefactor $\CK_{\CO_1\CO_2}$ of a three-point function and then their action on $t^{\CO_1\CO_2}_{\CO_3}$ can be fully recast as a derivative with respect to $X _3,\Theta_3,\Thetab_3$.
\par
Firstly we need to compute the exact linear combinations of descendants that appear in the $\ell\ellb\neq 0$ terms. This has been done already in~\cite{Li:2014gpa} by analyzing superconformal two-point functions. We summarize their results in \appendixname~\ref{app:details}. Then we need an ansatz for the differential operator that we wish to compute. Schematically we have
\eqn{
\DQ{} \sim D_\alpha\,,\qquad \DQb{} \sim \Db_\alphad\,.
}[]
Therefore an ansatz for $\DQlQblb$ will be something of the form
\eqn{
\DQlQblb \sim a_1 \, (\DQ{})^\ell(\DQb{})^\ellb + \mbox{ permutations}\,,
}[]
and the coefficients $a_1,\ldots$ need to be fixed in terms of the $c_i$'s in (\ref{thetathetab}--\ref{thetasqsq}). This matching could be done by simply working out the algebra of the chiral derivatives, namely $\{D_\alpha,\Db_\alphad\} = 2\sigma^\mu_{\alpha\alphad}\,\partial_\mu$. However, we opted for a more convenient method. The strategy is to define a functional that acts on the non-supersymmetric operators $O(\bfx)$ and turns them into an explicit function $\varphi[O](\bfx)$. It is then possible to implement the rules for derivatives and index contractions in a computer algebra system and impose the following equality.
\eqn{
\varphi\big[\DQlQblb\,\CO(\bfz)\big|_0\big](\bfx) = \varphi\big[(Q^\ell\Qb{}^\ellb O)\big](\bfx)\,.
}[functionalphi]
The functional $\varphi$ can be chosen arbitrarily as long as it is generic enough to make \eqref{functionalphi} imply \eqref{generalDllbdef}.\footnote{Another way to say this is the following: we want to prove some identities between differential operators. The identities should hold for any choice of functions to which the operators are applied. Therefore we need to find a set of functions which are generic enough to completely fix the ansatz, but also as easy as possible to manipulate.} There are a few advantages to this method. First, it can be easily implemented using the package introduced in \sectionname~\ref{sec:mathematica}. Second, it is possible to choose among many functionals thus obtaining an overconstrained system of equations like \eqref{functionalphi}. The existence of a solution serves as a check for our results.
\par
A possible choice is $\varphi[O](\bfx) = \langle X O(\bfx) \rangle$. However if $X$ is a local operator then $\varphi$ becomes a quadratic functional (because necessarily $X = \Ob$) and it is hard to solve the constraints. If $X$ is the product of two local operators we have a linear functional but there is no choice that gives a nonzero three-point function with all possible $O$s. Naturally, there is no reason why $\varphi$ needs to be a physical correlator. This is then our choice:
\eqn{
\varphi[O](\bfx) = \frac{(\chi\llsp\rmx\llsp\etab)^l (\eta\llsp\rmx \chib)^k}{x^{\Delta_O+l+k}}\,(\eta\chi)^{j-k}\,(\etab\chib)^{\jb-l}\,.
}[]
The parameters $k,l$ can be varied between $0$ and, respectively, $j$ and $\jb$ to obtain a family of functionals. The only identity that needs to be considered when comparing derivatives of the above expression is the following
\eqn{
\chi\llsp\rmx\llsp\chib\; \eta\llsp\rmx\llsp\etab = \chi\llsp\rmx\llsp\etab\;\eta\llsp\rmx\llsp\chib + x^2\,\eta\chi\;\etab\chib\,.
}[eq:shout]
This is particularly convenient because the main obstacle in solving \eqref{functionalphi} is finding all linearly independent tensors. But, thanks to \eqref{eq:shout}, a basis of independent tensors can be simply taken to be
\eqn{
(\chi\llsp\rmx\llsp\chib)^n (\eta\llsp\rmx \etab)^m (\chi\llsp\rmx\llsp\etab)^a (\eta\llsp\rmx \chib)^b \,(\eta\chi)^c\,(\etab\chib)^d\,,\qquad \mbox{with $m \lsp n = 0$}\,.
}[]
Now the task of fixing the ansatz for the differential operators $\DQlQblb$ is tedious but entirely straightforward.

\subsubsection{First order}

At first order in $\theta$ and $\thetab$ no descendants need to be subtracted. The differential operators are simply $\partial_\theta$ or $\partial_\thetab$, which can be then completed to chiral derivatives~\eqref{chiralDDbdef}
\eqna{
\DQp &= - \frac{i}{j+1}\,\eta^\alpha D_\alpha\,,\qquad &
\DQm &= - \frac{i}{j}\,\frac{\partial}{\partial \eta_\alpha} D_\alpha\,,\\
\DQbp &= - \frac{i}{\jb+1}\,\etab^\alphad \Db_\alphad\,,\qquad &
\DQbm &= - \frac{i}{\jb}\,\frac{\partial}{\partial \etab_\alphad} \Db_\alphad\,.\\
}[opFirst]
As needed, these operators satisfy
\eqn{
\DQ\pm\,\CO(\bfz)\big|_0 = (QO)^\pm(\bfx)\,,\qquad
\DQb\pm\,\CO(\bfz)\big|_0 = (\Qb O)^\pm(\bfx)\,.
}[]
They will be used as building blocks for the subsequent differential operators.

\subsubsection[Orders \texorpdfstring{$Q^2$ and $\Qb{}^2$}{Q^2 and Qb^2}]{Orders $\boldsymbol{Q^2}$ and $\boldsymbol{\Qb{}^2}$}

For the quadratic order we may use the operators $D^\alpha D_\alpha$ and $\Db_\alphad\Db^\alphad$. However in the next section we want to prove that all $\DQlQblb$ commute with the prefactor of the three-point function. This can be done easily only if all operators are expressed in terms of $\DQ\pm$ and $\DQb\pm$. Since we act on homogeneous functions of $\eta$ and $\etab$ this amounts to only an overall factor.\footnote{
Indeed for an homogeneous function $f_\ell = \eta_{\alpha_1}\cdots \eta_{\alpha_\ell}\,f^{\alpha_1\ldots\alpha_\ell}$ one has
\[
\begin{aligned}
\partial_\eta D\,\eta D\,f_\ell &= \tfrac12 \lsp D^2\lsp \partial_{\eta_\beta}\, \eta_\beta\eta_{\alpha_1}\cdots \eta_{\alpha_\ell}\,f^{\alpha_1\ldots\alpha_\ell} \\
& = \tfrac12(\ell + 2)\lsp D^2\, f_\ell\,.
\end{aligned}
\]
}
The result is
\eqn{
\DQsq = \frac{2\lsp j(j+1)}{j+2}\,\DQm\,\DQp\,,\qquad
\DQbsq = -\frac{2\lsp \jb(\jb+1)}{\jb+2}\,\DQbm\,\DQbp\,.
}[opDsq]
The factors $j(j+1)$ in the numerator simply cancel the denominator of \eqref{opFirst}. For scalar operators they can be omitted and $\DQm$ is used without the $j$ at the denominator.

\subsubsection[Order \texorpdfstring{$Q\Qb$}{Q Qb}]{Order $\boldsymbol{Q\Qb}$}

This is the first order where we need to subtract the descendants. Schematically we have
\eqn{
\CO\big|_{\theta\thetab} = \theta\thetab\left((Q\Qb O) - i c\,\partial_\mu O\right)\,,
}[]
see \eqref{thetathetab} for the full expression. The needed ansatz is simple. Letting $s$ and $r$ represent either a plus or a minus sign we have
\eqn{
\DQQb{s}{r} = a^{sr}\,\DQ{s}\DQb{r} + b^{sr}\,\DQb{r}\DQ{s}\,.
}[DQQbdef]
The coefficients $a^{sr}$ and $b^{sr}$ are a function of $c_1$ if $sr=++$, $c_2$ if $sr = -+$, $c_3$ if $sr = +-$ and $c_4$ if $sr=--$. We will directly give the final expression by replacing the values of $c_i$ computed in~\cite{Li:2014gpa}.
\eqna{
a^{++} &= \frac{2q+j}{2(q+\qb)+j+\jb}\,,\qquad &
b^{++} &= -\frac{2\qb+\jb}{2(q+\qb)+j+\jb}\,,\\
a^{-+} &= \frac{2q-j-2}{2(q+\qb-1)-j+\jb}\,,\qquad &
b^{-+} &= -\frac{2\qb+\jb}{2(q+\qb-1)-j+\jb}\,,\\
a^{+-} &= \frac{2q+j}{2(q+\qb-1)+j-\jb}\,,\qquad &
b^{+-} &= -\frac{2\qb-\jb-2}{2(q+\qb-1)+j-\jb}\,,\\
a^{--} &= \frac{2q-j-2}{2(q+\qb-2)-j-\jb}\,,\qquad &
b^{--} &= -\frac{2\qb-\jb-2}{2(q+\qb-2)-j-\jb}\,.\\
}[abdef]
We should remark that these expressions are valid for a generic long multiplet. When the multiplet is short (e.g. $2q = j+2$) the differential operators associated to null superdescendants should be discarded.
\par
As a quick example we can take the $Q\Qb$ descendant of the Ferrara-Zumino multiplet $J(\bfz)$, namely the energy-momentum tensor. The result is simply one half the commutator
\eqn{
T(\bfx) =  \frac12\,\big(\DQp\DQbp - \DQbp\DQp\big)\,J(\bfz)\big|_0\,,
}[]
as can be easily seen by letting $j=\jb=1$ and $q=\qb=3/2$.

\subsubsection[Orders \texorpdfstring{$Q^2\Qb$ and $\Qb{}^2 Q$}{Q^2 Qb and Qb^2 Q}]{Orders $\boldsymbol{Q^2\Qb}$ and $\boldsymbol{\Qb{}^2 Q}$}

The contribution at this order is shown in equation~\eqref{thetasqthetab}. Letting $s = \pm$, the ansatz is
\eqna{
\DQbsqQ{s} &= c^s\,\DQ{s}\,\DQbm\,\DQbp + d^s\,\DQbm\,\DQ{s}\,\DQbp + e^s\,\,\DQbm\,\DQbp\,\DQ{s}\,, \\
\DQsqQb{s} &= \bar{c}^s\,\DQb{s}\,\DQm\,\DQp + \bar{d}^s\,\DQm\,\DQb{s}\,\DQp + \bar{e}^s\,\,\DQm\,\DQp\,\DQb{s}\,. \\
}[Dorderthree]
The various coefficients are a function of $c_{5,6,7,8}$ and $\bar{c}_{5,6,7,8}$. We have the following simple relation 
\eqn{
c^s,\,d^s,\,e^s = - \lsp\bar{c}^s ,\,- \lsp\bar{d}^s ,\,- \lsp\bar{e}^s
\,\big |{}_{j\leftrightarrow \jb,\,q\leftrightarrow \qb}\,,\quad
}[]
It will then suffice to quote the result for the coefficients of $\DQbsqQ{\pm}$ only
\eqnal{
c^+ &= \frac{2 \jb (\jb+1) (2 q+j) (j+\jb+2 q+2 \qb+2)}{(\jb+2) (\jb-j-2 q-2 \qb+2) (j+\jb+2 q+2 \qb)}\,, \\
c^- &= \frac{2 \jb (\jb+1) (2 q-j-2) (\jb-j+2 q+2 \qb)}{(\jb+2) (j+\jb-2 q-2 \qb+4) (\jb-j+2 q+2 \qb-2)}\,,\\
d^+ & = \frac{4 \jb (\jb+1) (2 q+j)}{(\jb-j-2 q-2 \qb+2) (j+\jb+2 q+2 \qb)}\,, \\
d^- &=  \frac{4 \jb (\jb+1) (2 q-j-2)}{(j+\jb-2 q-2 \qb+4) (\jb-j+2 q+2 \qb-2)}
\\
e^+ &=-\frac{2 \jb (\jb+1) (2 \qb+\jb)}{(\jb+2) (j+\jb+2 q+2 \qb)} \,,\\
e^- &=-\frac{2 \jb (\jb+1) (2 \qb+\jb)}{(\jb+2) (\jb-j+2 q+2 \qb-2)}\,.
}[cdedef]

\subsubsection[Order \texorpdfstring{$Q^2\Qb{}^2$}{Q^2 Qb^2}]{Order $\boldsymbol{Q^2\Qb{}^2}$}

At last we have the highest order in $\theta$ and $\thetab$. The subtractions needed are six: $c_9$ through $c_{14}$. This means that our ansatz will need seven terms obtained by permuting $\DQp,\DQm,\DQbp$ and $\DQbm$. In total there are fourteen permutations after taking into account $\{D,D\} = \{\Db,\Db\} = 0$. Not all of these are independent and the choice of seven out of these is not unique. We made this ansatz
\eqna{
\DQsqQbsq = & \phantom{\,+\,}
f_1\,\DQbm\,\DQbp\,\DQm\,\DQp+f_2\,\DQbm\,\DQm\,\DQbp\,\DQp
+f_3\,\DQbm\,\DQm\,\DQp\,\DQbp\\
&+f_4\,\DQm\,\DQbm\,\DQbp\,\DQp+f_5\,\DQm\,\DQbm\,\DQp\,\DQbp\\
&+f_6\,\DQm\,\DQp\,\DQbm\,\DQbp+f_7\,\DQp\,\DQbm\,\DQbp\,\DQm\,.
}[]
These are the values of the coefficients $f_i$
\eqnal{
f_1 &=\frac{4 j (j+1) \jb (\jb+1) (2 \qb-\jb-2) (\jb+2 \qb) (j+\jb+2 q+2 \qb+2)}{(j+2) (\jb+2) (j-\jb-2 q-2 \qb+2) (j-\jb+2 q+2 \qb-2) (j+\jb+2 q+2 \qb)}\,,\\
f_2 &=\frac{16 j (j+1) \jb (\jb+1) (2 \qb-\jb-2) (2 \qb+\jb)}{(j-\jb-2 q-2 \qb+2) (j+\jb-2 q-2 \qb+4) (j-\jb+2 q+2 \qb-2) (j+\jb+2 q+2 \qb)}\,,\\
f_3 &=-\frac{8 j (j+1) \jb (\jb+1) (2 q+j)}{(j+2)}\,\times\\
&\phantom{=\;}\times
\frac{\left(j^2+4 j \qb+2 j+\jb^2+2 \jb-4 q^2-8 q \qb+4 q-4 \qb^2+12 \qb\right)}{(j-\jb-2 q-2 \qb+2) (j+\jb-2 q-2 \qb+4) (j-\jb+2 q+2 \qb-2) (j+\jb+2 q+2 \qb)}\,,\\
f_4 &=\frac{4 j \jb (\jb+1) (2 q-j-2) (2 \qb+\jb) (j-\jb-2 q-2 \qb)}{(\jb+2) (j-\jb-2 q-2 \qb+2) (j+\jb-2 q-2 \qb+4) (j+\jb+2 q+2 \qb)}\,,\\
f_5 &=\frac{16 j (j+1) \jb (\jb+1) (2 q-j-2) (2q+j)}{(j-\jb-2 q-2 \qb+2) (j+\jb-2 q-2 \qb+4) (j-\jb+2 q+2 \qb-2) (j+\jb+2 q+2 \qb)}\,,\\
f_6 &=\frac{4 j (j+1) \jb (\jb+1) (2 q-j-2) (2 q+j) (j+\jb+2 q+2 \qb+2)}{(j+2) (\jb+2) (j-\jb-2 q-2 \qb+2) (j-\jb+2 q+2 \qb-2) (j+\jb+2 q+2 \qb)}\,,\\
f_7 &=\frac{4 j \jb (\jb+1) (2q+j) (2\qb+\jb) (j+\jb+2 q+2 \qb+2)}{(\jb+2) (\jb-j+2 q+2 \qb-2) (j-\jb+2 q+2 \qb-2) (j+\jb+2 q+2 \qb)}\,.
}[fidef]
All these expressions are available within the Mathematica package that we introduce in \sectionname~\ref{sec:mathematica}.

\subsection[\texorpdfstring{$\CN=2$}{N = 2} case]{$\boldsymbol{\CN=2}$ case}
\subsubsection{General remarks}

\begin{figure}[tbp]
\centering
\includegraphics[scale=1.18]{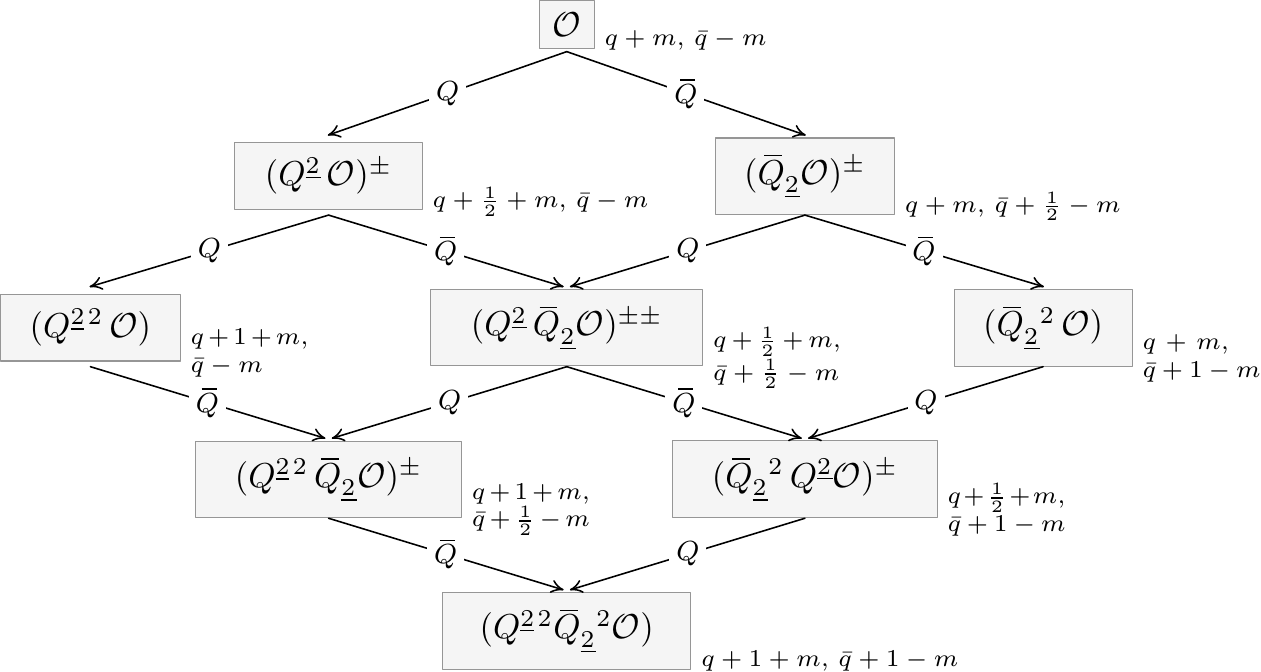}
\caption{Diagram of all $\CN=1$ multiplets in an $\CN=2$ long multiplet. Superscripts of $\pm$ indicate the choice of spin $j\pm1$ or $\jb\pm1$. Subscripts indicate the $\CN=1$ $q,\lsp\qb$ charges. Here $m = R_3$ takes integer spaced values between $-R/2$ and $R/2$. The $\mathfrak{u}(1)_{\CN=2}$ R-charge grows from left to right by $1$ and the conformal dimension grows from top to bottom by $1/2$. A box represents a family of long $\CN=1$ multiplets.}
\label{fig:N2}
\end{figure}

Now we want to define differential operators that extract full $\CN=1$ superconformal multiplets inside an $\CN=2$ multiplet. We have chosen $Q^{\un1}$ and $\Qb_{\un1}$ as generators of the subalgebra. The decomposition is illustrated schematically in \figurename~\ref{fig:N2}. We define $(Q^{\un2\lsp \ell}\Qb_{\un2}{}^{\lsp\ellb}\CO)$ to be a family of $\CN=1$ superconformal primaries, with a slight abuse of notation. In fact, due to the anticommutation relations\footnote{The generators $R^I_{\phantom{I}\lsp J}$ for $I \neq J$ are the $\mathfrak{su}(2)_R$ ladder operators $R_\pm$.}
\eqn{
\{ Q^I_\alpha,\, S_J^\beta\} = -4 \lsp\delta_\alpha^\beta\, R^I_{\phantom{I}\lsp J}\,,\qquad
\{ \overbar{S}^{I\alphad},\, \Qb_{J\betad}\} = -4 \lsp\delta_\alphad^\betad\, R^I_{\phantom{I}\lsp J}\,,
\;\qquad I \neq J\,,
}[QScommrel]
in general such operators will not be obtained by simply acting with $Q^{\un2}$ and $\Qb_{\un2}$ on $\CO$ and thus some subtractions might be needed. Each of these families will decompose in $R+1$ multiplets with the following charges
\eqn{
\big(q+m,\,\qb  - m\lsp\big)\,,\qquad
m = -\tfrac12R,\,-\tfrac12R+1,\ldots,\tfrac12R\,,
}[RchargesList]
where we denoted the $\CN=2$ charges with $q,\qb$ and the $\mathfrak{su}(2)$ representation of $\CO$ with $R$. In this paper we only consider operators with $R = \{0,1,2\}$ and give results up to the first nontrivial order in the Grassmann variables with $I=2$, leaving the general analysis for a future work.

\subsubsection{Zero R-charge}

If the superconformal primary is an $\mathfrak{su}(2)$ singlet, no computation is needed at the linear order. Indeed from \eqref{QScommrel} and from the fact that $R_\pm$ on a singlet yields zero, one can see that these operators are automatically $\CN=1$ superconformal primaries
\eqn{
{}_{\un2}\DQ\pm\,\CO(\bfz)\big|_0 = (Q^{\un2}O)^\pm(\bfz)\,,\qquad
{}_{\un2}\DQb\pm\,\CO(\bfz)\big|_0 = (\Qb_{\un2} O)^\pm(\bfz)\,,
}[]
where now we defined $|_0$ as $|_{\theta_{\un2} = \theta^{\un2}=0}$ and the prefix on the differential operator  as an $\mathfrak{su}(2)$ index
\eqn{
{}_I\lnsp\DQlQblb \equiv \DQlQblb \raisebox{-.5ex}{$\big|_{D \to D^I,\, \Db \to \Db_I}$}\,.
}[IDdefinition]
Similarly, the order, $Q^2$ and $\Qb{}^2$ require no subtractions as well and can be defined by attaching an index $I=2$ to \eqref{opDsq}.
\par
More interesting is the order $Q\Qb$. We expect a single superconformal primary at this level. The correct differential operator is a linear combination of the operators in \eqref{DQQbdef} for $I=1$ and $2$. We will not prove this result here but postpone the discussion to \sectionname~\ref{sec:Rchargezero}. Let us denote with a boldface $\BBD$ the $\CN=2$ differential operators. Letting $s$ and $t$ represent a sign $\pm$, the result is
\eqn{
\BBD_{Q\Qb}^{st} = {}_{\un2}\DQQb{s}{t} + A^{st}\; {}_{\un1}\DQQb{s}{t}\,,
}[opBBDdef]
with
\eqna{
A^{++} &= -\frac{2}{2( q+\qb+1)+j+\jb}\,,&\qquad
A^{-+} &= -\frac{2}{2( q+\qb)-j+\jb}\,,\\
A^{--} &= -\frac{2}{2(q+\qb-1) - j -\jb}\,,&\qquad
A^{+-} &= -\frac{2}{2( q+\qb)+j-\jb}\,.
}[Asol]
As a quick example we can reproduce the result of \cite{Kuzenko:1999pi} for the stress tensor multiplet. Let us denote with $\CJ(z)$ the $\CN=2$ superconformal primary and with $J(\bfz)$ the Ferrara-Zumino multiplet. Recalling that for $\CJ$ we have $q = \qb = 1$ and $j = \jb = 0$, the result is
\eqn{
J(\bfz) = \BBD_{Q\Qb}^{++}\,\CJ \big|_{0}= \left({}_{\un2}\DQQbpp\,\CJ  - \frac13\;{}_{\un1}\DQQbpp\,\CJ \right)\big|_{0}\,,
}[]
with, from \eqref{DQQbdef} and \eqref{opFirst},
\eqn{
 {}_{I}\DQQbpp\,\CJ = -\frac12\lsp[D_\alpha^I,\,\Db_{\alphad I}]\lsp\CJ\,.
}[]
Apart from an overall minus sign, which simply reflects in a different normalization, we obtain the same linear combination.

\subsubsection[R-charge \texorpdfstring{$1/2$}{1/2}]{R-charge $\boldsymbol{1/2}$}

When the $\mathfrak{su}(2)$ R-charge is non-zero, the differential operators are nontrivial starting from the first order. The simplest case is that of the doublet, for which we expect two $\CN=1$ superconformal primaries. The operator $\CO$ on which they act will have an $I$ index. Letting $s = \pm$, we can write general ansatze as
\eqn{
{}_J\BBD_Q^s \,\CO_J= {}_I\DQ{s}\;\CM_s^{IJ} \,\CO_J \,,\qquad
{}_J\BBD_{\Qb}^s \,\CO_J= {}_I\DQb{s}\;\overbar{\CM}_s^{IJ} \,\CO_J\,.
}[ansatzChargehalf]
We denoted with $\BBD$ the $\CN=2$ differential operator. In order to have a superconformally covariant operator we need to contract the index of $\CO$ with an appropriate matrix. Since we expect two multiplets to arise at each order, the solutions for $\CM_s$ and $\overbar{\CM}_s$ must have two degrees of freedom each. We denote the two classes of solutions as $\CA$ and $\CB$. The matrix $\CM$ may be an arbitrary linear combination of those solutions, but the basis that we chose is the one that projects into $\CN=1$ multiplets of definite R-charge. In order to avoid confusion we indicate here what term is represented by each entry
\eqna{
\CM_s = \left(
\begin{matrix}
{}_{\un1}\DQ{s}\,\CO_{\un1} &\quad {}_{\un1}\DQ{s}\,\CO_{\un2} \\
{}_{\un2}\DQ{s}\,\CO_{\un1}&\quad {}_{\un2}\DQ{s}\,\CO_{\un2}
\end{matrix}
\right)\,,\qquad\mbox{similarly for $\overbar{\CM}_s$}\,.
}[]
The $\CA$ solution for $\CM$ will be $\CM_s = \CA_s$, $\overbar{\CM}_s = \overbar{\CA}_s$, with
\eqn{
\CA_s = \left(
\begin{array}{cc}
0 & 0 \\ 1 & 0
\end{array}
\right)\,,\qquad
\overbar{\CA}_s = \left(
\begin{array}{cc}
0 & 0 \\ 0 & 1
\end{array}
\right)\,.
}[Asolution]
The solutions does not depend on $s$. Whereas the $\CB$ solution for $\CM$ will be $\CM_s = \CB_s$, $\overbar{\CM}_s = \overbar{\CB}_s$, with 
\eqna{
\CB_+ &= \left(
\begin{array}{cc}
-\frac{2}{2q+j+1} & 0 \\ 0 & 1
\end{array}
\right)\,,&\qquad
\CB_- &= \left(
\begin{array}{cc}
-\frac{2}{2q-j-1} & 0 \\ 0 & 1
\end{array}
\right)\,,\\
\overbar{\CB}_+ &= \left(
\begin{array}{cc}
0& \frac{2}{2\qb+\jb+1} \\ 1 & 0
\end{array}
\right)\,,&\qquad
\overbar{\CB}_- &= \left(
\begin{array}{cc}
0& \frac{2}{2\qb-\jb-1} \\ 1 & 0
\end{array}
\right)\,.
}[Bsolution]
In this case a nontrivial linear combination is needed to obtain an $\CN=1$ superconformal primary and the solution does depend on the sign $s$. The resulting operators will have charges and spin dictated by \figurename~\ref{fig:N2}. As before, we defer the proof of these results to \sectionname~\ref{sec:Rchargehalf}.

\subsubsection[R-charge \texorpdfstring{$1$}{1}]{R-charge $\boldsymbol{1}$}

The case of $\mathfrak{su}(2)$ R-charge $1$ is not conceptually different from the last section. Now the operator will have an adjoint index $A$ and the matrices $\CM$ in \eqref{ansatzChargehalf} will be rectangular.
\eqn{
{}_J\BBD_Q^s \,\CO_J= {}_I\DQ{s}\;\CM_s^{IA} \,\CO_A \,,\qquad
{}_J\BBD_{\Qb}^s \,\CO_J= {}_I\DQb{s}\;\overbar{\CM}_s^{IA} \,\CO_A\,.
}[ansatzChargeone]
We expect three degrees of freedom for the choice of $\CM_s$ and $\overbar{\CM}_s$. Thus we can span the basis by three classes of solutions $\CA$, $\CB$ and $\CC$. For the reader's convenience we will show all of them at once by taking an arbitrary linear combination of them\footnote{As before, this choice of basis is not arbitrary but it is the one that projects on $\CN=1$ multiplets with definite R-charges.}
\eqna{
a\, \CA_+ + b\, \CB_+ + c\,\CC_+ &= \left(
\begin{array}{ccc}
\frac{2\lsp c}{2q+j+2} & \frac{2i\lsp c}{2q+j+2} & - \frac{2i \lsp b}{2q+j} \\
\frac12(a+i\llsp b) & \frac{i}2(a-i\llsp b) & c
\end{array}
\right)\,,\\
a\, \CA_- + b\, \CB_- + c\,\CC_- &= \left(
\begin{array}{ccc}
\frac{2\lsp c}{2q-j} & \frac{2i\lsp c}{2q-j} & - \frac{2i \lsp b}{2q-j-2} \\
\frac12(a+i\llsp b) & \frac{i}2(a-i\llsp b) & c
\end{array}
\right)\,,\\
a\, \overbar{\CA}_+ + b\, \overbar{\CB}_+ + c\,\overbarUp{\CC}_+ &= \left(
\begin{array}{ccc}
\frac{2\lsp c}{2\qb+\jb+2} & -\frac{2i\lsp c}{2\qb+\jb+2} & \frac{2i \lsp b}{2\qb+\jb} \\
\frac12(a-i\llsp b) & -\frac{i}2(a+i\llsp b) & c
\end{array}
\right)\,,\\
a\, \overbar{\CA}_- + b\, \overbar{\CB}_- + c\,\overbarUp{\CC}_- &= \left(
\begin{array}{ccc}
\frac{2\lsp c}{2\qb-\jb} & -\frac{2i\lsp c}{2\qb-\jb} & \frac{2\lsp b}{2\qb-\jb-2} \\
\frac12(a-i\llsp b) & -\frac{i}2(a+i\llsp b) & c
\end{array}
\right)\,.
}[ABCsolution]
As before, the various $\CN=1$ superconformal primaries are obtained by \eqref{ansatzChargeone} by replacing $\CM$ with either $\CA,\lsp\CB$ or $\CC$. The proof these results is postponed to \sectionname~\ref{sec:Rchargeone}.

\section{Acting on three-point functions}\label{sec:threepf}

\subsection[\texorpdfstring{$\CN=1$}{N = 1} case]{$\boldsymbol{\CN=1}$ case}
\subsubsection{Idea}

The main goal is to fix a basis of non supersymmetric three-point functions for a given triplet of representations
$
\langle \prod_{i=1}^3\,O_i(\bfx_i)\rangle^{(a)}
$, where $a = 1,\ldots n_{123}$,
and to expand the three-point function of a superdescendant in that basis. Namely we want to find the coefficients $\lambda^{(a)}$ such that
\eqn{
\DQlQblb \,\langle
 \mbox{$\prod_{i=1}^3\,$}\CO_i(\bfz_i)
\rangle\big|_0 = \sum_{a=1}^{n_{(1^{\ell\ellb})23}} \lambda^{(a)}\,\langle(Q^\ell\Qb{}^\ellb O_1)(\bfx_1)O_2(\bfx_2)O_3(\bfx_3)\rangle^{(a)}\,,
}[comparehard]
provided that the full superconformal three-point function is known.\footnote{We implicitly defined $n_{(1^{\ell\ellb})23}$ as the number of tensor structures in $\langle(Q^\ell\Qb{}^\ellb O_1)O_2O_3\rangle$.} A three-point function can be decomposed as \eqref{GeneralThreepf}. When we act on it with, say, $\DQp$ we get two terms. But since $\CK_{\CO_1\CO_2}$ is bosonic, when the Grassmann variables are set to zero only one survives
\eqn{
\DQp \langle \mbox{$\prod_{i=1}^3\,$}\CO_i(\bfz_i)\rangle\big|_0 = \CK_{\CO_1\CO_2}\;\DQp\,t^{\CO_1\CO_2}_{\CO_3}(Z_3)\big|_0\,.
}[firststep]
This is certainly convenient as we do not have to worry about the prefactor when taking derivatives, but it is not yet what we need. If we want to compare the above expression with a chosen basis of three-point functions we still need to expand the definitions of $Z_3$ and to act with the spinor derivatives inside $\CK_{\CO_1\CO_2}$. It would be much better if we could express \eqref{firststep} as
\eqn{
\DQp \langle \mbox{$\prod_{i=1}^3\,$}\CO_i(\bfz_i)\rangle\big|_0 = \CK_{(Q\CO_1)^+\CO_2}\;\CDQbp\,t^{\CO_1\CO_2}_{\CO_3}(Z_3)\big|_0\,,
}[secondstep]
following \cite{Kuzenko:1999pi}. Here $\CK_{(Q\CO_1)^+\CO_2}$ is the prefactor of an hypothetical three-point function of $(Q\CO_1)^+,\,\CO_2$ and $\CO_3$, if $(Q\CO_1)^+$ were a superconformal primary. It is simply a $\CK_{\CO_1\CO_2}$ with shifted arguments
\eqna{
\CK_{(Q\CO_1)^\pm \CO_2} = \CK_{\CO_1\CO_2}\raisebox{-2ex}{\bigg |}{}_{\substack{q_1\to q_1-1/2\\\qb_1\to \qb_1+1\hspace{1.7ex}\\j_1\to j_1\pm1\hspace{1.7ex}}}
\,,\qquad
\CK_{(\Qb\CO_1)^\pm \CO_2} = \CK_{\CO_1\CO_2}\raisebox{-2ex}{\bigg |}{}_{\substack{\qb_1\to \qb_1-1/2\\q_1\to q_1\pm1\hspace{1.7ex}\\\jb_1\to \jb_1+1\hspace{1.7ex}}}\,.
}[shiftsQ]
And $\CDQbp$ will be defined later together with all the details, but the important point is that it is a differential operator with respect to the variables $X_3,\Theta_3$ and $\Thetab_3$. Now the problem is drastically simplified. We can choose a basis of non-nilpotent tensor structures in $t^{\CO_1\CO_2}_{\CO_3}$, 
\eqn{
t^{\CO_1\CO_2}_{\CO_3; \,a}(X_3)\,,\qquad a = 1,\ldots,n_{123}\,,
}[]
which in turn will induce a basis of non-supersymmetric three-point functions. And then the comparison can be done at the level of the $t$. Assuming for now that this reasoning works for all superdescendants one has\footnote{Notice the swap $\ell \leftrightarrow \ellb$.}
\eqn{
\CDQlbQbl \lsp\,t^{\CO_1\CO_2}_{\CO_3}(Z_3)\big|_0 = \sum_{a=1}^{n_{(1^{\ell\ellb})23}} \lambda^{(a)}\,t^{(Q^\ell\Qb{}^\ellb \CO_1)\CO_2}_{\CO_3; \,a}(X_3)\,,
}[compareasy]
in place of \eqref{comparehard}. It is evident that \eqref{compareasy} is easier to solve for $\lambda^{(a)}$. But we went too fast in all the steps involved. First we need to show that $\DQlQblb$ actually commutes with the prefactor for any $\ell, \ellb$. Then we also need to prove that \eqref{secondstep} is always possible and define the $\CDQlbQbl$ operators that arise from it.
\par
The proof of \eqref{secondstep} is not hard. We need to make use of the formulas \eqref{DonF}, which are valid for any function of $Z_3 = X_3,\Theta_3,\Thetab_3$. The extra factors of $\rmx_{i\jb}$ that appear can be reabsorbed in the prefactor and they give automatically the right shifts in the quantum numbers. Then one can define the derivatives
\eqna{
\CDQp &= \frac{1}{\jb+1}\,\chi^\alpha \CD_\alpha\,,\qquad &
\CDQm &= -\frac{1}{\jb}\,\frac{\partial}{\partial \chi_\alpha} \CD_\alpha\,,\\
\CDQbp &= \frac{1}{j+1}\,\chib^\alphad \overbar{\CD}_\alphad\,,\qquad &
\CDQbm &= -\frac{1}{j}\,\frac{\partial}{\partial \chib_\alphad} \overbar{\CD}_\alphad\,,\\
}[CopFirst]
in complete analogy with \eqref{opFirst}.\footnote{Notice the swap $j\leftrightarrow\jb$.} The detailed expressions are given in \appendixname~\ref{app:identities}. Clearly one can also define, by repeated application, the following operators
\eqn{
\CDQlbQbl = \DQlQblb\big|_{D\to\overbar{\CD},\,\, \Db \to \CD}\,.
}[CDdefinition]
Checking for commutativity with $\CK_{\CO_1\CO_2}$ requires us to show
\eqn{
\left(\DQlQblb \,\CK_{\CO_1\CO_2} \,t^{\CO_1\CO_2}_{\CO_3} - \CK_{(Q^\ell\Qb{}^\ellb)\CO_1\CO_2} \,\CDQlbQbl \,t^{\CO_1\CO_2}_{\CO_3}\right)\big|_0=0\,.
}[commutativity]
This is trivially true if $\ell \ellb$ vanishes: If $\ell$ or $\ellb$ is $1$ then we can do the same argument as the example of before: the derivative acting on $\CK_{\CO_1\CO_2}$ is necessarily fermionic and thus vanishing if the Grassman variables are set to zero. When, on the other hand, $\ell$ or $\ellb$ is $2$, there are two pieces. One is fermionic an thus vanishing and the other must be proportional to $\theta^2$ or $\thetab^2$ due to its R-charge scaling. If $\ell\ellb \neq 0$ the result is non-trivial and will be proven explicitly in the next paragraphs. We can however argue that the commutativity property must hold without any computation. Indeed it is easy to convince oneself that the terms in \eqref{commutativity} that survive after setting the $\theta$'s to zero cannot recombine to form an expression with the right prefactor and a function of $X_3$. Therefore, if they did not vanish, the result of $\DQlQblb$ applied on a correlator would not be a three-point function of conformal primaries but of a combination of primaries and descendants. This is a contradiction by construction of the operator $\DQlQblb$. We will nevertheless carry an explicit computation in order to have a non-trivial check of our results.
The rest of this section will be devoted to show that \eqref{commutativity} holds and thus complete the proof of \eqref{compareasy}.

\subsubsection[Order \texorpdfstring{$Q\Qb$}{Q Qb}]{Order $\boldsymbol{Q\Qb}$}\label{sec:commutativityQQb}

We want to show \eqref{commutativity} for $\ell=\ellb=1$. There are two kinds of terms: those where one derivative acts on $\CK$ and one on $t$ and those where both act on $\CK$. After setting the Grassmann variables to zero only the latter may survive, so we need to focus on them. Concretely we need to show
\eqn{
(a^{sr}\,\DQ{s}\DQb{r} + b^{sr}\,\DQb{r}\DQ{s})\,\lsp\CK_{\CO_1\CO_2} = 0\,.
}[]
We now use \eqref{DonK} for the first derivative and \eqref{DonKtheta} for the second one. The expressions obtained for different values of $s,r = \pm$ will be proportional to the following factors:
\eqna{
s=r=1 \;:\qquad & a^{++}\, (2\qb+\jb) + b^{++}\,(2q + j)\,,\\
-s=r=1\;: \qquad & a^{-+}\, (2\qb+\jb) + b^{-+}\,(2q - j -2)\,,\\
s=-r=1\;: \qquad & a^{+-}\, (2\qb-\jb-2) + b^{+-}\,(2q + j)\,,\\
s=r=-1\;: \qquad & a^{--}\, (2\qb-\jb-2) + b^{--}\,(2q - j -2)\,.
}[]
From \eqref{abdef} it is easy to verify that all these quantities are zero and thus the derivative commutes with the prefactor as needed. As we commented earlier, this depends crucially on the fact that the differential operators do not yield conformal descendants.

\subsubsection[Orders \texorpdfstring{$Q^2\Qb$ and $\Qb{}^2 Q$}{Q^2 Qb and Qb^2 Q}]{Orders $\boldsymbol{Q^2\Qb}$ and $\boldsymbol{\Qb{}^2 Q}$}

For this order, only the terms with one derivative on $t$ and two derivatives on $\CK$ can survive. Furthermore, the derivatives on $\CK$ must be with respect to $Q$ and $\Qb$. Since also applying derivatives on $t$ shifts the quantum numbers (see \eqref{KDont}), one needs to be careful with the ordering. For the derivatives that act on $\CK$ we use \eqref{DonK} and \eqref{DonKtheta} as before. There are in total eight different cases (see \eqref{Dorderthree}): in $\DQsqQb{s}$ either the $\DQp$ or the $\DQm$ may act on the $t$ and $s$ may be $\pm$. Similarly in $\DQbsqQ{s}$, $\DQbp$ or $\DQbm$ may act on the $t$ and $s$ may be $\pm$. For brevity we only illustrate two cases. If $\DQbp$ acts on the $t$ and $s = +$, the result is proportional to
\eqn{
(2q+j)\,e^+ - (2q+j+2)\,d^+ - (2\qb-\jb-4)\,c^+ \,.
}[]
If, on the other hand, $\DQm$ acts on the $t$ and $c=-$, the result is proportional to
\eqn{
(2\qb-\jb-2)\,\bar{e}^- - (2\qb-\jb)\,\bar{d}^- - (2q-j-4)\,\bar{c}^- \,.
}[]
In all cases it can be checked that the resulting expressions vanish when one replaces the coefficients with \eqref{cdedef}.

\subsubsection[Order \texorpdfstring{$Q^2\Qb{}^2$}{Q^2 Qb^2}]{Order $\boldsymbol{Q^2\Qb{}^2}$}\label{sec:commutativityQsqQbsq}

This is the last and most challenging order. The terms that can survive are of two kinds: those where two derivatives ($\DQ{}$ and $\DQb{}$) act on the $t$ and the other two on the $\CK$ and those where all the derivatives act on $\CK$. Working out these cases in the same way as we did before requires deriving formulas for repeated applications of the differential operators, similar to those appearing in \appendixname~\ref{app:identities}. We preferred resorting to a ``brute force'' approach instead. We used the explicit definition of $\CK$ and applied the derivatives on it using the Mathematica package introduced in \sectionname~\ref{sec:mathematica}. The case where all derivatives act on $\CK$ is straightforward and can be done with the functions defined in the package. The case where only two derivatives act on $\CK$ requires a small explanation first.
Since $t$ is a generic function, we cannot take explicit derivatives of it. But we can always modify \eqref{DonF} as follows
\eqna{
\,D_{1\lsp\alpha} \,t(Z_3) &= -i \frac{(\rmx_{1\bar3})_{\alpha\alphad}}{{x_{\bar13}}^2}\,\bar{\xi}^\alphad\;\overset{\leftarrow}{\partial}_{\bar{\xi}}\partial_{\bar{\xi}'}\;\bar{\xi}'\overbar{\CD}\,t(Z_3)\,,\\
\,\Db_{1\lsp\alphad} \,t(Z_3) &= -i \lsp\xi^\alpha\frac{(\rmx_{3\bar1})_{\alpha\alphad}}{{x_{\bar31}}^2}\;\partial_{\xi'}\overset{\leftarrow}{\partial}_{\xi}\;\xi'\CD\,t(Z_3)\,,
}[factorizet]
where $\xi,\xi',\bar{\xi},\bar{\xi}'$ are other auxiliary polarization. In this way we can factorize either a $\xi'\CD\,\bar{\xi}'\overbar{\CD}t$ or a $\bar{\xi}'\overbar{\CD}\,\xi'\CD t$ and focus on the rest. Now the problem becomes explicit and one can check whether the resulting expressions vanish.
\par
We performed this computation and observed that, with the values of $f_i$ given in \eqref{fidef}, all expressions identically vanish. This completes the proof of \eqref{commutativity}.

\subsection[\texorpdfstring{$\CN=2$}{N = 2} case]{$\boldsymbol{\CN=2}$ case}
\subsubsection{Lowest order}

The lowest order $\theta_{\un2} = \thetab^{\un2} = 0$ is almost entirely trivial. The matrices $\hat{u}_I^{\phantom{I}\lsp J}$ appearing in the prefactor of \eqref{GeneralThreePFN2} reduce to
\eqn{
\hat{u}_1^{\phantom{1}1}(z_{13})\big|_{\theta_{\un2} = \thetab^{\un2} = 0} = \left( \frac{{x_{\bar13}}^2}{{x_{\bar31}}^2}\right)^{\frac12}\,,\qquad
\hat{u}_2^{\phantom{2}2}(z_{13})\big|_{\theta_{\un2} = \thetab^{\un2} = 0} = \left( \frac{{x_{\bar31}}^2}{{x_{\bar13}}^2}\right)^{\frac12}\,,
}[uhat1122def]
the off-diagonal components being zero. It is also obvious from \eqref{xijdef} that all $\CN=2$ quantities that depend on $x_{\ib j}$ simply reduce to the same quantity but with the $\CN=1$ definition of $x_{\ib j}$. By looking at the prefactor \eqref{Kdef} it is easy to see that the factors of $\hat{u}$ can be absorbed by shifting the $q,\qb$ labels as follows
\eqn{
\hat{u}_1^{\phantom{1}1}(z_{13})\,\CK_{\CO_1\CO_2} =  \CK_{\CO_1\CO_2}\raisebox{-1.3ex}{\Big |}{}_{\substack{q_1\to q_1+1/2\\\qb_1\to \qb_1-1/2}}\,,\qquad
\hat{u}_2^{\phantom{2}2}(z_{13})\,\CK_{\CO_1\CO_2} =  \CK_{\CO_1\CO_2}\raisebox{-1.3ex}{\Big |}{}_{\substack{q_1\to q_1-1/2\\\qb_1\to \qb_1+1/2}}\,.
}[shiftsR]
The component of $\CO$ with $\mathfrak{su}(2)$ R-charge $R_3= m$ will have a prefactor containing
$
(\hat{u}_1^{\phantom{1}\lsp1})^{\frac12 R-m}$ $(\hat{u}_2^{\phantom{2}\lsp2})^{\frac12R + m}
$. This contributes to an $\CN=1$ superconformal primary with $q,\qb$ charges equal to $(q - m, \qb + m)$, consistently with \eqref{RchargesList}.

\subsubsection{Zero R-charge}\label{sec:Rchargezero}

If the superconformal primary is an $\mathfrak{su}(2)$ singlet, the commutativity with the prefactor follows immediately from the $\CN=1$ case. Indeed the structure of $\CK_{\CO_1\CO_2}$ is identical except for the fact that $\rmx_{i\jb}$ has more Grassmann variables. The crucial observation is that only $\rmx_{3\ib}$ or $\rmx_{i\bar3}$ appear. We can thus write
\eqna{
\rmx_{i\bar3} =\big(\rmx_{i3} &- 2i \lsp\theta_{\un1 i}\,\thetab^{\un1}_{i} - 2i \lsp\theta_{\un1 3}\,\thetab^{\un1}_{3} + 4i\lsp \theta_{\un1 i}\,\thetab^{\un1}_{3}\,\big)\\
&  - 2i \lsp\theta_{\un2 i}\,\thetab^{\un2}_{i} - 2i \lsp\theta_{\un2 3}\,\thetab^{\un2}_{3} + 4i\lsp \theta_{\un2 i}\,\thetab^{\un2}_{3}\,,
}[]
Since there is no term mixing $\theta_{\un1}$ and $\theta_{\un2}$ we can simply rename the quantity inside the parentheses as $\rmx'_{i3}$ and carry the same exact computation as the $\CN=1$ case. The same argument applies to $\rmx_{3\ib}$. The necessary shifts that need to be applied to the prefactor differ sightly from \eqref{shiftsQ}. They follow directly from \eqref{DonFN2}
\eqna{
\CK_{(Q^I\CO_1)^\pm \CO_2} &= \hat{u}_J^{\phantom{J}I}(z_{31})\,\CK_{\CO_1\CO_2}\raisebox{-1.3ex}{\Big |}{}_{\substack{\qb_1\to \qb_1+1/2\\j_1\to j_1\pm1/2}}
\,,\\
\CK_{(\Qb_I\CO_1)^\pm \CO_2} &= \hat{u}_I^{\phantom{I}J}(z_{13})\,\CK_{\CO_1\CO_2}\raisebox{-1.3ex}{\Big |}{}_{\substack{q_1\to q_1+1/2\\\jb_1\to \jb_1\pm1/2}}\,.
}[shiftsQN2]
Then, using \eqref{DonFN2} followed by setting the $\theta_{\un2}, \thetab^{\un2}$ Grassmann variables to zero results in an $\CN=1$ superconformal correlator. The identities presented in \eqref{shiftsR}, \eqref{shiftsQN2} imply that the resulting superconformal primary has the desired $q,\qb$ charges: $(q + 1/2,\,\qb)$ for the $Q^{\un2}$ descendant and $(q,\,\qb + 1/2)$ for the $\Qb_{\un2}$ descendant.\footnote{To see this, one needs to use the property $\hat{u}_2^{\phantom22}(z_{31}) = \hat{u}_1^{\phantom11}(z_{13})
$.\label{note}
} The result is consistent with \figurename~\ref{fig:N2}.
\par
At order $Q^{\un2}\Qb_{\un2}$ instead we need to use the operator $\BBD_{Q\Qb}$ defined in \eqref{opBBDdef}. Here we will adopt a different strategy from the $\CN=1$ case. We will actually use the commutativity with the prefactor to derive the form of the differential operator. The reason why this is a valid proof is that, thanks to the formulas \eqref{DonFN2}, we can show that the action of such an operator on a three-point function yields a correlator of a superconformal primary. We could have followed the same approach for the $\CN=1$ case, of course. However in the way we did it the prefactor commutativity served as an important nontrivial check of our results. The computation is a bit more involved than that of \sectionname~\ref{sec:commutativityQQb} because we are not setting all $\theta$'s to zero but only $\theta_{\un2}$ and $\thetab^{\un2}$. In particular, there are non vanishing contributions also from terms where only one derivative acts on $\CK$. We follow a ``brute force'' approach similar to that of \sectionname~\ref{sec:commutativityQsqQbsq}: we act either with both differential operators $D$ and $\Db$ on the $\CK$ or we act with one of them on the $t$ and we factorize it away using \eqref{factorizet}. Letting $|_0 \equiv |_{\theta_{\un2} = \theta^{\un2}=0}$, the following equation has a unique solution for $A^{st}$, given by \eqref{Asol}:
\begin{multline}
\big({}_{\un2}\DQQb{s}{t} + A^{st}\; {}_{\un1}\DQQb{s}{t}\big)\,\CK_{\CO_1\CO_2}  \,t^{\CO_1\CO_2}_{\CO_3}\,\big|_0 =
\\
\hfill= \CK_{(Q^{\un2}\Qb_{\un2}\CO_1)\CO_2}\,\hat{u}_{2}^{\phantom22}(z_{13})\lsp\hat{u}_{1}^{\phantom11}(z_{13})\,\big({}_{\un2}\CDQQb{s}{t} + A^{st}\; {}_{\un1}\CDQQb{s}{t}\big)t_{\CO_3}^{\CO_1\CO_2}\,\big|_0\,,
\end{multline}
where we used the property in footnote~\ref{note}. In analogy with \eqref{IDdefinition} and \eqref{CDdefinition} we defined
\eqn{
{}_I\CDQlbQbl \equiv \CDQlbQbl \raisebox{-.5ex}{$\big|_{\CD \to \CD_I,\, \overbar{\CD} \to \overbar{\CD}^I}$}\,.
}[]
According to the shifts defined in \eqref{shiftsQN2} and the definitions in \eqref{uhat1122def} the result of the action on the $t$ is a superconformal primary with charges $(q+1/2,\qb+1/2)$ as expected from \figurename~\ref{fig:N2}. We can then define the analog of the $\BBD_{Q\Qb}$ when acting on the $t$ as
\eqn{
\mathfrak{D}_{Q\Qb}^{st} = {}_{\un2}\CDQQb{s}{t} + A^{st}\; {}_{\un1}\CDQQb{s}{t}\,.
}[mathfrakDdef]
 
\subsubsection[R-charge \texorpdfstring{$1/2$}{1/2}]{R-charge $\boldsymbol{1/2}$}\label{sec:Rchargehalf}

Following an approach similar to the last subsection here we will claim that \eqref{ansatzChargehalf} annihilates the prefactor. The difference now is that the prefactor has indices: $\CK_{\CO_1\CO_2}\hat{u}_I^{\phantom{I}J}$ one of which is contracted with the matrix $\CM$. By explicitly computing the action of the differential operators on the prefactor we can impose that it vanishes and use this to fix the matrix $\CM$. We discover that there are two possible solutions for each case: $\CA$ defined in \eqref{Asolution} and $\CB$ defined in \eqref{Bsolution}. These two choices will give rise to two independent $\CN=1$ multiplets when acting on the $t$. They will have charges $(q+1/2\pm1/2,\qb\mp1/2)$ for the $Q$ descendant and $(q\pm1/2,\qb+1/2\mp1/2)$ for the $\Qb$ descendant, as described by \figurename~\ref{fig:N2}. If we choose $\CM = \CA$ we have (a $|_0$ is implicit in all the following formulas)
\eqna{
{}_I\DQ{s}\,\CA_s^{IJ}\,\CK_{\CO_1\CO_2}\lsp \hat{u}_J^{\phantom{J}K}(z_{13})\lsp t_{\CO_3}^{\CO_1\CO_2 \lsp| K} &= \CK_{(Q^{\un2}\CO_1)\CO_2}\,\hat{u}_1^{\phantom11}(z_{13})\lsp\hat{u}_2^{\phantom22}(z_{31})\,\lsp{}_{\un2}\CDQb{s}\lsp t_{\CO_3}^{\CO_1\CO_2 \lsp|\lsp 1}\,,\\
{}_I\DQb{s}\,\overbar{\CA}_s^{IJ}\,\CK_{\CO_1\CO_2}\lsp \hat{u}_J^{\phantom{J}K}(z_{13})\lsp t_{\CO_3}^{\CO_1\CO_2\lsp| K} &= \CK_{(\Qb_{\un2}\CO_1)\CO_2}\,\hat{u}_2^{\phantom22}(z_{13})\lsp\hat{u}_2^{\phantom22}(z_{13})\,\lsp{}_{\un2}\CDQ{s}\lsp t_{\CO_3}^{\CO_1\CO_2\lsp|\lsp2}\,.
}[]
The first line corresponds to the $Q^{\un2}$ descendant with charges $(q+1,\qb-1/2)$, while the second line corresponds to the $\Qb_{\un2}$ descendant with charges $(q-1/2,\qb+1)$, as can be seen from \eqref{shiftsR}, \eqref{shiftsQN2} and the property in footnote~\ref{note}. Similarly, if we choose $\CM= \CB$ we have
\eqna{
{}_I\DQ{s}\,\CB_s^{IJ}\,\CK_{\CO_1\CO_2}\lsp \hat{u}_J^{\phantom{J}K}(z_{13})\lsp t_{\CO_3}^{\CO_1\CO_2 \lsp | K} &= \CK_{(Q^{\un2}\CO_1)\CO_2}\,\hat{u}_1^{\phantom11}(z_{13})\lsp\hat{u}_2^{\phantom22}(z_{13})\,\times\\
&\phantom{=}\;\times\left(\lsp{}_{\un2}\CDQb{s}\, t_{\CO_3}^{\CO_1\CO_2 \lsp | \lsp2 } + \CB_s^{11}\,\lsp{}_{\un1}\CDQb{s}\, t_{\CO_3}^{\CO_1\CO_2 \lsp | \lsp1 }\lsp\right)\,,\\
{}_I\DQb{s}\,\overbar{\CB}_s^{IJ}\,\CK_{\CO_1\CO_2}\lsp \hat{u}_J^{\phantom{J}K}(z_{13})\lsp t_{\CO_3}^{\CO_1\CO_2 \lsp | K} &= \CK_{(\Qb_{\un2}\CO_1)\CO_2}\,\hat{u}_1^{\phantom11}(z_{13})\lsp\hat{u}_2^{\phantom22}(z_{13})\,\times\\
&\phantom{=}\;\times\left(\lsp{}_{\un2}\CDQ{s}\, t_{\CO_3}^{\CO_1\CO_2 \lsp | \lsp1 } + \overbar{\CB}_s^{12}\,\lsp{}_{\un1}\CDQ{s}\, t_{\CO_3}^{\CO_1\CO_2 \lsp | \lsp2 }\lsp\right)\,,
}[]
where again we have used the property of footnote~\ref{note}. Now the first line corresponds to the $Q^{\un2}$ descendant with charges $(q,\qb+1/2)$ and the second line corresponds to the $\Qb_{\un2}$ descendant with charges $(q+1/2,\qb)$.

\subsubsection[R-charge \texorpdfstring{$1$}{1}]{R-charge $\boldsymbol{1}$}\label{sec:Rchargeone}

There are no qualitative differences between the cases with $R=1$ and $R=2$. As before we act on the prefactor, impose that it vanishes, and solve for the matrix $\CM$. This will give rise to three different choices, $\CA,\CB$ and $\CC$ defined in \eqref{ABCsolution}. We will now list all possible ways of acting on the $t$ and show the charges of the $\CN=1$ superconformal primaries that are produced. In order to streamline the notation we will denote with $\CT \equiv \CT^{\mathbf2}$ the R-charge~$1$ ($R=2$) representation that appears in \eqref{ThalfTone}. We will also define
\eqn{
t_{\CO_3}^{\CO_1\CO_2 \lsp | \lsp \pm} = \frac12\left(t_{\CO_3}^{\CO_1\CO_2 \lsp | \lsp 1} \pm i \lsp t_{\CO_3}^{\CO_1\CO_2 \lsp | \lsp 2}\right)\,.
}[]
Let us start from $\CA$
\eqna{
{}_I\DQ{s}\,\CA_s^{IA}\,\CK_{\CO_1\CO_2}\lsp \CT(\hat{u}(z_{13}))_A^{\phantom{A}B}\lsp t_{\CO_3}^{\CO_1\CO_2 \lsp | B} &= \CK_{(Q^{\un2}\CO_1)\CO_2}\,(\hat{u}_1^{\phantom11}(z_{13}))^3\,\lsp{}_{\un2}\CDQb{s}\, t_{\CO_3}^{\CO_1\CO_2 \lsp | \lsp+ }\,,\\
{}_I\DQb{s}\,\overbar{\CA}_s^{IA}\,\CK_{\CO_1\CO_2}\lsp \CT(\hat{u}(z_{13}))_A^{\phantom{A}B}\lsp t_{\CO_3}^{\CO_1\CO_2 \lsp | B} &= \CK_{(\Qb_{\un2}\CO_1)\CO_2}\,(\hat{u}_2^{\phantom22}(z_{13}))^3\,\lsp{}_{\un2}\CDQ{s}\, t_{\CO_3}^{\CO_1\CO_2 \lsp | \lsp- }\,.
}[]
These represent $\CN=1$ superconformal primaries with charges, respectively, $(q+3/2,\qb-1)$ and $(q-1,\qb+3/2)$. Then we continue with $\CB$
\eqna{
{}_I\DQ{s}\,\CB_s^{IA}\,\CK_{\CO_1\CO_2}\lsp \CT(\hat{u}(z_{13}))_A^{\phantom{A}B}\lsp t_{\CO_3}^{\CO_1\CO_2 \lsp | B} &= \CK_{(Q^{\un2}\CO_1)\CO_2}\,\hat{u}_2^{\phantom22}(z_{13})\,\times\\&\phantom{=}\;\times i \lsp\left(\lsp{}_{\un2}\CDQb{s}\, t_{\CO_3}^{\CO_1\CO_2 \lsp | \lsp-} + \CB_s^{13}\, {}_{\un1}\CDQb{s}\, t_{\CO_3}^{\CO_1\CO_2 \lsp | \lsp3}\right)\,,\\
{}_I\DQb{s}\,\overbar{\CB}_s^{IA}\,\CK_{\CO_1\CO_2}\lsp \CT(\hat{u}(z_{13}))_A^{\phantom{A}B}\lsp t_{\CO_3}^{\CO_1\CO_2 \lsp | B} &= \CK_{(\Qb_{\un2}\CO_1)\CO_2}\,\hat{u}_1^{\phantom11}(z_{13})\,\times\\&\phantom{=}\;\times (-i)\left(\lsp{}_{\un2}\CDQ{s}\, t_{\CO_3}^{\CO_1\CO_2 \lsp | \lsp+} + \overbar{\CB}_s^{13}\, {}_{\un1}\CDQ{s}\, t_{\CO_3}^{\CO_1\CO_2 \lsp | \lsp3}\right)\,.
}[]
These represent $\CN=1$ superconformal primaries with charges, respectively, $(q-1/2,\qb+1)$ and $(q+1,\qb-1/2)$. Finally we have $\CC$
\eqna{
{}_I\DQ{s}\,\CC_s^{IA}\,\CK_{\CO_1\CO_2}\lsp \CT(\hat{u}(z_{13}))_A^{\phantom{A}B}\lsp t_{\CO_3}^{\CO_1\CO_2 \lsp | B} &= \CK_{(Q^{\un2}\CO_1)\CO_2}\,\hat{u}_1^{\phantom11}(z_{13})\,\times\\&\phantom{=}\;\times \lsp\left(\lsp{}_{\un2}\CDQb{s}\, t_{\CO_3}^{\CO_1\CO_2 \lsp | \lsp3} + 2\lsp\CC_s^{11}\, {}_{\un1}\CDQb{s}\, t_{\CO_3}^{\CO_1\CO_2 \lsp | \lsp+}\right)\,,\\
{}_I\DQb{s}\,\overbar{\CC}_s^{IA}\,\CK_{\CO_1\CO_2}\lsp \CT(\hat{u}(z_{13}))_A^{\phantom{A}B}\lsp t_{\CO_3}^{\CO_1\CO_2 \lsp | B} &= \CK_{(\Qb_{\un2}\CO_1)\CO_2}\,\hat{u}_2^{\phantom22}(z_{13})\,\times\\&\phantom{=}\;\times \left(\lsp{}_{\un2}\CDQ{s}\, t_{\CO_3}^{\CO_1\CO_2 \lsp | \lsp3} + 2\lsp\overbarUp{\CC}_s^{11}\, {}_{\un1}\CDQ{s}\, t_{\CO_3}^{\CO_1\CO_2 \lsp | \lsp-}\right)\,.
}[]
Here we also used $\CC^{11} = -i \lsp\CC^{12}$ and $\overbarUp{\CC}^{11} = i\lsp\overbarUp{\CC}^{12}$. These last operators represent $\CN=1$ superconformal primaries with charges, respectively, $(q+1/2,\qb)$ and $(q,\qb+1/2)$.

\section{Exotic chiral primaries}
\label{sec:exotic}

The chiral operators are $\CN=2$ superconformal multiplets that satisfy an $L\bar{B}_1$ shortening condition.\footnote{In the notation of~\cite{Dolan:2002zh} this shortening condition is denoted as $\overbar{\CE}$.} That means that they are annihilated by the $\Qb$ supercharges
\eqn{
[\lsp\Qb_{I\alphad}\,,\,\CX(\bfz)\lsp \} = 0\,,\qquad I=1,2\,.
}
This dictates that they must have spin $(j,0)$, with $j$ any non-negative integer, and their conformal dimension must be half their $\mathfrak{u}(1)$ R-charge $r$. In terms of the $q,\,\qb$ charges in \eqref{qqbcharges} one has $q = r/2$ and $\qb = 0$. The exotic primaries $\CX$ are defined as those chiral operators with $j>0$ that are $\mathfrak{su}(2)$ singlets. If $j=0$ the chiral operators are often called Coulomb branch operators. There exist also chiral operators with non vanishing $\mathfrak{su}(2)$ R-charge but we will not consider those here. The statement that we will prove is that the exotic primaries cannot appear in any local SCFT, that is, any SCFT which admits a conserved traceless stress tensor.
\par 
We start by constructing a three-point function of $\CX$, its conjugate\footnote{The conjugate has R-charge $-r$, spin $(0,j)$ and charges $q = 0,\,\qb = r/2$. It satisfies a $B_1\bar{L}$ shortening.} $\CXb$ and the stress tensor multiplet $\CJ$. Let us choose the following parametrization
\eqn{
\langle \CXb(\bfz_1)\lsp\CJ(z_2)\lsp\CX(\bfz_3)\rangle = \CK_{\CXb\CJ}\,t^{\CXb\CJ}_{\CX}(Z_3,\chi_1,\eta_3)\,.
}[]
The function $t^{\CXb\CJ}_{\CX}$ must be chiral at point $z_1$, for example. This implies
\eqn{
\overbar{\CD}^I_\alphad\,\lsp t^{\CXb\CJ}_{\CX}(Z_3,\chi_1,\eta_3) = 0\,,\qquad I=1,2\,.
}[]
By using the representation \eqref{CDdefX} of the differential operators one can see that $t^{\CXb\CJ}_{\CX}$ may only depend on $X_3$ and $\Theta_3^I$. But since the sum of R-charges is zero there cannot be an isolated $\Theta_3^I$ and thus $t^{\CXb\CJ}_{\CX}$ is a function of $X_3$, which can be fixed by scaling \eqref{scalingt} up to an overall constant.
\eqn{
t^{\CXb\CJ}_{\CX}(Z_3,\eta_1,\eta_3) = \CA \lsp \frac{(\eta_1\llsp\eta_3)^{\llsp j}}{{X_3}^2}\,.
}[exotic3pf]
It is also easy to verify that the conservation of $\CJ$ is satisfied. The conservation operators are $(\CQ_I)^2$ and $(\overbar{\CQ}^I)^2$. If we rewrite \eqref{exotic3pf} as a function of $\Xb_3,\,\Theta_{3I},\,\Thetab{}^I_3$ we see that there are no terms of order $\Theta_3^2\lsp \Thetab{}^2_3$. This means that we can use the representation \eqref{CQdefThetab} for $(\overbar{\CQ}^I)^2$ and the representation \eqref{CQdefTheta} for $(\CQ_I)^2$ and get trivially zero in both cases.
\par
The proof now consists in showing that \eqref{exotic3pf} does not satisfy the stress tensor Ward identities unless $j=0$. In order to obtain the Ward identities we have to expand this function in components and extract the contribution from $T(\bfx_2)$, the stress tensor, and $J(\bfx_2)$, the R-symmetry current. Then impose integral equalities of the form
\twoseqn{
\int_\Sigma \di\Omega^\mu \,\varepsilon^\nu\lsp \langle\CXb(\bfx_1)\lsp T_{\mu\nu}(x_2)\lsp\CX(\bfx_3)\rangle &= (\Delta + x_3\cdot\partial_3)\,\langle \CXb(\bfx_1)\lsp \CX(\bfx_3) \rangle\,.
}[WIdilat]{
\int_\Sigma \di\Omega^\mu \,\langle\CXb(\bfx_1)\lsp J_{\mu}(x_2)\lsp\CX(\bfx_3)\rangle &= r\,\langle \CXb(\bfx_1)\lsp \CX(\bfx_3) \rangle\,.
}[][]
Here $\Sigma$ is a three-dimensional surface enclosing $x_2$ and $x_3$ but not $x_1$, $\di\Omega^\mu$ is its volume element and $\varepsilon^\mu$ is the dilatation conformal Killing vector. An identity similar to \eqref{WIdilat} follows from the translation Killing vector. Fortunately, however, half of the work has already been done. It suffices to extract the contribution from the Ferrara-Zumino multiplet $J(\bfz_2)$ inside of $\CJ(\bfz_2)$. Then we can use the results of~\cite{Manenti:2019kbl} to apply the Ward identities.
\par
We will keep denoting as $\CX$ the $\CN=1$ primary appearing as the lowest order in the exotic operator multiplet. Extracting the contribution of the Ferrara-Zumino gives
\eqn{
\langle \CXb(\bfz_1)\lsp J(\bfz_2)\lsp\CX(\bfz_3)\rangle = \CK_{\CXb(Q^{\un2}\Qb_{\un2}\CJ)}\,\mathfrak{Q}_{Q\Qb}^{++}\,t^{\CXb\CJ}_{\CX}(Z_3,\chi_1,\eta_3)\lsp\big|_{0}\,,
}[]
with the definition $\mathfrak{Q}_{Q\Qb}^{++} = \mathfrak{D}_{Q\Qb}^{++}\big|_{\CD \to \CQ}$, $\mathfrak{D}$ defined in \eqref{mathfrakDdef} and $\CQ$ defined in \eqref{CQdef}. The result is
\eqn{
\mathfrak{Q}_{Q\Qb}^{++}\,t^{\CXb\CJ}_{\CX}(Z_3,\eta_1,\eta_3)\lsp\big|_{0} = - \frac{4\llsp i}{3} \CA\,(\eta_1\eta_3)^j \lsp\frac{\eta_2\llsp X_3\llsp\etab_2}{{X_3}^4}\,,
}[mathfrakQt]
where now $X_3$ follows the $\CN=1$ definition. A three-point function of an $\CN=1$ chiral operator instead reads\footnote{In the basis of \cite{Manenti:2019kbl} the coefficients $A_{1,2}$ are translated into the coefficients $\CC_{1,\ldots,10}$ as follows:\[
\CC_1 = \tfrac12\lsp \CC_4 = - \tfrac14\lsp \CC_5 = A_1+A_2,\,\qquad\CC_2 = \tfrac12\lsp \CC_7 = -\tfrac14\lsp \CC_8 = -A_2\,,\qquad \CC_{3,6,9,10}=0\,.
\]}
\eqn{
t^{\CXb J}_{\CX}(Z_3,\eta_1,\eta_3)\lsp = i\lsp A_1 \,(\eta_1\eta_3)^{\llsp j} \lsp\frac{\eta_2\llsp X_3\llsp\etab_2}{{X_3}^4} + i\lsp A_2\,(\eta_1\eta_3)^{\llsp j-1} \lsp\frac{\eta_1\lnsp X_3\llsp\etab_2\,\eta_2\eta_3}{{X_3}^4}\,.
}[N1chiral3pf]
The comparison between \eqref{mathfrakQt} and \eqref{N1chiral3pf} is straightforward and yields
\eqn{
A_1 = - \frac{4}{3}\lsp\CA\,,\qquad A_2 = 0\,,
}[]
while the Ward identities for chiral operators require
\eqn{
A_1 = i^{\llsp j}\,\frac{r - 3\llsp j}{3\llsp \pi^2}\,,\qquad A_2 = i^{\llsp j}\,\frac{2\llsp j}{\pi^2}\,.
}[]
This immediately implies that there are no solutions for $j > 0$ and thus the exotic primaries cannot couple consistently with the stress tensor and must be absent from any local theory. It also tells us that if $j = 0$ then $\CA$ is fixed to be
\eqn{
\CA = -\frac{r}{4\llsp\pi^2}\,.
}[CAr4pi]
As a check of our formalism we expanded \eqref{exotic3pf} to higher orders in the supercharges and extracted also the contributions from $Q^{\un2}\lsp\CX$ and $Q^{\un2\lsp2}\CX$, which are all $\CN=1$ $L\bar{B}_1$ chiral multiplets. We verified that for $j=0$ all the components satisfy the Ward identities when $\CA$ takes the value in \eqref{CAr4pi}.

\section{A \texorpdfstring{\textit{Mathematica}}{Mathematica} package}\label{sec:mathematica}

Computations in superspace, in particular those necessary to solve \eqref{compareasy}, might be hard to do by hand. We introduce a Mathematica package as a convenient tool to perform such tasks. It can be found in the repository \href{\mathematicapack}{\texttt{gitlab.com/maneandrea/spinoralgebra}}. There is also a version of this package that only deals with commuting variables, which can be used for any tensor computation in four dimensions. A complete documentation is made available in the form of a notebook.\par
The package works with the index-free formalism following the same conventions as this paper. There are different input and output notations available: one for improved readability and one to write code more easily. It is possible to reduce, Taylor expand and compare expressions. Furthermore, many differential operators have been defined, including the chiral derivatives (\ref{chiralDDbdef}, \ref{CDdef}, \ref{CQdef}) and all the operators appearing in (\ref{lowest}\,--\,\ref{thetasqsq}). It contains a precomputed two-point function for general values of $q,\qb,j,\jb$. By including the package \href{https://gitlab.com/bootstrapcollaboration/CFTs4D}{\texttt{CFTs4D}}~\cite{Cuomo:2017wme} it is also possible to use the functionalities for $\CN=1$ superspace three-point functions. For any three given operators, the package gives a basis of $t^{\CO_1\CO_2}_{\CO_3}$ tensor structures.
\renewcommand{\ttdefault}{DejaVuSansMono-TLF}
\subsection{Note on the conventions}\label{sec:packconvention}
For two-point function we follow the conventions of~\cite{Li:2014gpa}. Namely if a superprimary $\CO$ has a two-point function given by
\eqn{
\langle\COb(\bfz_1)\CO(\bfz_2)\rangle = C_\CO \frac{(\eta_1\rmx_{1\bar2}\etab_2)^j(\eta_2\rmx_{\bar12}\etab_1)^\jb}{{x_{1\bar2}}^{2q+j}{x_{\bar12}}^{2\qb+\jb}}\,,\qquad (-i)^{j+\jb}C_\CO > 0\,,
}[]
then any of its descendants will have a non-supersymmetric two-point function given by
\eqn{
\langle\Ob{}'(\bfx_1)\,O'(\bfx_2)\rangle = C_{(Q^\ell\Qb{}^\ellb O)} \frac{(\eta_1\rmx_{12}\etab_2)^{j'}(\eta_2\rmx_{12}\etab_1)^{\jb{}'}}{{x_{12}}^{2(q'+\qb')+j'+\jb{}'}}\,,\qquad O'(\bfx) \equiv (Q^\ell\Qb{}^\ellb O)(\bfx)\,,
}[]
with the ratio of the respective normalizations fixed. From the package, it can be obtained as follows
\eqn{
\frac{C_{(Q^\ell\Qb{}^\ellb O)}}{C_\CO} = \texttt{\small operatorNorm["}Q^\ell\Qb{}^\ellb O\texttt{\small", \{q,qb\}, \{j,jb\}]}\,.
}[]
Supersymmetric three-point functions instead can be computed by the function \texttt{\footnotesize SUSY3pf}. The expressions are given directly in the space where $t_{\CO_3}^{\CO_1\CO_2}(X,\Theta,\Thetab)$ lives, which we will call ``$t$ space'' in this paragraph. The notation of the package is as follows
\eqn{
\texttt{\small x3} \to X_3\,,\qquad \texttt{\small\straighttheta3} \to \Theta_3\,,\qquad \texttt{\small\straighttheta b3} \to \Thetab_3\,.
}
The structures are generated by calling internally the package \texttt{\footnotesize CFTs4D} and then following the procedure explained in \sectionname~\ref{sec:N1superspace}. Naturally one can work in $t$ space also for non-supersymmetric three-point functions using the non nilpotent supersymmetric structures as a basis. Expanding any function of $X_3,\eta_i,\etab_i$ in such a basis does not require much computational effort. It may however be useful to make contact with more familiar bases. In \tablename~\ref{tab:mapping} we show the mapping between the $\Theta,\Thetab\to0$ limit of the non-nilpotent structures in $t$ space and the embedding space structures in \texttt{\footnotesize CFTs4D} \cite{Cuomo:2017wme}. Every tensor structure can be constructed as a monomial over the building blocks listed in \tablename~\ref{tab:mapping}. In order to pass from $t$ space to embedding one has to multiply both sides by the appropriate prefactors. The structures in $t$ space need to be multiplied by $\CK_{\CO_1\CO_2}$ times an overall scaling $X_3^a$. While the embedding structures should be multiplied by the kinematic prefactor given in \texttt{\footnotesize CFTs4D} as \texttt{\footnotesize{n3KinematicFactor}}. In formulas one has
\eqn{
\CK_{\CO_1\CO_2}\,X_3^{2(\Delta_3-\Delta_2-\Delta_1)}\big|_0\cdot\big(\mbox{$t$ space}\big) = \prod_{\substack{i<j\\k\neq i,j}}^3 |x_{ij}|^{\kappa_k-\kappa_i-\kappa_j}\cdot \big(\mbox{embedding}\big)\,,
}[eq:mapping]
with
\eqn{
\Delta_i \equiv q_i+\qb_i\,,\qquad \kappa_i \equiv \Delta_i + \tfrac12(j_i+\jb_i)\,.
}

\newcommand{\Ionetwo}{
\tableeqn{\phantom{-}\frac{\eta_1\lnsp X_3\etab_2}{(X_3^2)^{1\lnsp/2}}} & \tableeqn{\hat{\BBI}^{1,2}}
}
\newcommand{\Itwoone}{\tableeqn{-\frac{\eta_2\lnsp X_3\etab_1}{(X_3^2)^{1\lnsp/2}}} & \tableeqn{\hat{\BBI}^{2,1}}
}
\newcommand{\Kone}{\tableeqn{-\frac{\eta_3\lnsp X_3\etab_2}{(X_3^2)^{1\lnsp/2}}} & \tableeqn{\hat{\BBK}^{2,3}_{1}}
}
\newcommand{\Ionethree}{\tableeqn{\phantom{-}\eta_1\eta_3} & \tableeqn{\hat{\BBI}^{1,3}}
}
\newcommand{\Ithreeone}{\tableeqn{-\etab_1\etab_3} & \tableeqn{\hat{\BBI}^{3,1}} 
}
\newcommand{\Ktwo}{\tableeqn{-\frac{\eta_3\lnsp X_3\etab_1}{(X_3^2)^{1\lnsp/2}}} & \tableeqn{\hat{\BBK}^{3,1}_{2}}
}
\newcommand{\Kbone}{\tableeqn{-\frac{\eta_2\lnsp X_3\etab_3}{(X_3^2)^{1\lnsp/2}}} & \tableeqn{\hat{\overbarUp{\BBK}}^{2,3}_{1}}
}
\newcommand{\Kbtwo}{\tableeqn{-\frac{\eta_1\lnsp X_3\etab_3}{(X_3^2)^{1\lnsp/2}}} & \tableeqn{\hat{\overbarUp{\BBK}}^{3,1}_{2}}
}
\newcommand{\Kbthree}{\tableeqn{\phantom{-}\eta_1\eta_2} & \tableeqn{\hat{\overbarUp{\BBK}}^{1,2}_3}
}
\newcommand{\Kthree}{\tableeqn{\phantom{-}\etab_1\etab_2} & \tableeqn{\hat{\BBK}^{1,2}_3} 
}
\newcommand{\Jone}{\tableeqn{-\frac{\eta_1\lnsp X_3\etab_1}{(X_3^2)^{1\lnsp/2}}} & \tableeqn{\hat{\BBJ}^{1}_{2,3}}
}
\newcommand{\Jtwo}{\tableeqn{-\frac{\eta_2\lnsp X_3\etab_2}{(X_3^2)^{1\lnsp/2}}} & \tableeqn{\hat{\BBJ}^{2}_{3,1}}
}
\newcommand{\Jthree}{\tableeqn{-\frac{\eta_3\lnsp X_3\etab_3}{(X_3^2)^{1\lnsp/2}}} & \tableeqn{\hat{\BBJ}^{3}_{1,2}}
}
\newcommand{\Itwothree}{\tableeqn{\phantom{-}\eta_2\eta_3} & \tableeqn{\hat{\BBI}^{2,3}}
}
\newcommand{\Ithreetwo}{\tableeqn{-\etab_2\etab_3} & \tableeqn{\hat{\BBI}^{3,2}} 
}
\begin{table}[htbp]
\centering

\begin{tabular}[t]{cc}
$t$ space & Embedding \\
\hline
\\[-1em]
\Ionetwo
\\&\\[-1em]
\Itwoone 
\\&\\[-1em]
\Itwothree
\\&\\[-1em]
\Ithreetwo
\\&\\[-1em]
\Ionethree
\\&\\[-1em]
\Ithreeone
\\[.18em]\vdots&\vdots\\
\hline
\end{tabular}
\;
\begin{tabular}[t]{cc}
$t$ space & Embedding \\
\hline\\&\\[-2.8em]
\vdots&\vdots\\
\Kone
\\&\\[-1em]
\Ktwo
\\&\\[-1em]
\Kbone 
\\&\\[-1em]
\Kbtwo
\\[.9em]\vdots&\vdots\\
\hline
\end{tabular}
\;
\begin{tabular}[t]{cc}
$t$ space & Embedding \\
\hline\\&\\[-2.8em]
\vdots&\vdots\\
\\[-1.2em]
\Kthree
\\&\\[-1em]
\Kbthree
\\&\\[-1em]
\Jone
\\&\\[-1em]
\Jtwo 
\\&\\[-1em]
\Jthree\\
\\&\\[-1.95em]
\hline
\end{tabular}

\caption{Mapping between the $\Theta,\Thetab\to0$ limit of the non nilpotent three-point tensor structures in $t$ space and the embedding formalism structures in \cite{Cuomo:2017wme}. The equality between neighboring columns holds after we apply the appropriate $\CK_{\CO_1\CO_2} X_3^a$ prefactor to the $t$ space structures and the kinematic prefactor to the embedding structures. See \eqref{eq:mapping}.}\label{tab:mapping}
\end{table}

\subsection{Worked out example}
As an example, we illustrate step by step the computation of the chiral scalar blocks in the $\phi\times\phib$ OPE that appeared in \cite{Poland:2010wg}. We want to expand in $\theta_1,\,\thetab_1$ the correlator
\eqn{
\langle\CO(\bfz_1)\phi(z_2)\phib(z_3)\rangle = \CK_{\CO\phi}\,t^{\CO\phi}_\phib(Z_3,\chi_1,\chib_1)\,.
}[]
First we need to include the packages by calling \texttt{\footnotesize<\llsp<\,SpinorAlgebra\backtick} and \texttt{\footnotesize<\llsp<\,CFTs4D\backtick}. Then we generate the three-point function $t^{\CO\phi}_\phib$. In order to do that let us assume that the spin $j$ of $\CO$ is big enough 
\codein{
\$Assumptions = \{j > 20\};
}
\vspace{-2.5em}
\codein{
tO\straightphi\lsp\lnsp\straightphi b =  SUSY3pf[\{\{qO, qO\},\{q, 0\},\{0, q\}\},\{\{j, j\},\{0, 0\},\{0, 0\}\}]
}
\vspace{-1.5em}
\par\noindent
We also need to declare that $q_\CO$, $q$ and $j$ are constants. This is done by
\codein{
qO/:ConstQ[qO] := True;
}
\vspace{-1.5em}
\par\noindent
and similarly for the others. Now we can generate a basis of non-nilpotent tensor structures of the descendants. They are $(Q\Qb\CO)^{++},\,(Q\Qb\CO)^{--}$ and $(Q^2\Qb{}^2\CO)$. We can pass the \texttt{\footnotesize list} option to the function in order to output a basis and take the first part to obtain the non-nilpotent structures
\codein{
\raisebox{-1.3ex}{\parbox{10cm}{tQQbOpp = 
 SUSY3pf[\{\{qO+1/2,qO+1/2\},\{q, 0\},\{0,q\}\}, \{\{j+1,j+1\},\{0,0\},\{0,0\}\}, list $\to$ True][\!\![1]\!\!];}}}
 \vspace{-1em}
\codein{
\raisebox{-1.3ex}{\parbox{10cm}{tQQbOmm = 
 SUSY3pf[\{\{qO+1/2,qO+1/2\},\{q, 0\},\{0,q\}\}, \{\{j-1,j-1\},\{0,0\},\{0,0\}\}, list $\to$ True][\!\![1]\!\!];}}}
 \vspace{-1em}
 \codein{
\raisebox{-1.3ex}{\parbox{12cm}{tQsqQbsqO = 
 SUSY3pf[\{\{qO+1,qO+1\},\{q, 0\},\{0,q\}\}, \{\{j,j\},\{0,0\},\{0,0\}\}, list $\to$ True][\!\![1]\!\!];}}}
\vspace{-.5em}
\par\noindent
The results are in the package's notation but they can also be visualized in a clearer notation by wrapping them with the function \texttt{\footnotesize Nice}.
Now we can act with the differential operators on the superprimary and expand the result in the relative bases. For brevity we will only show the first case. Let us save in some variables the needed coefficients of \eqref{abdef}. They can be obtained from the package by calling the function \texttt{\footnotesize coefficientD}. For example
\codein{
\raisebox{-1.3ex}{
\parbox{9.2cm}{
app = coefficientD["a++", \{qO, qO\}, \{j, j\}];
bpp = coefficientD["b++", \{qO, qO\}, \{j, j\}];}}
}
\stepcounter{mathinput}
\vspace{-.5em}
\par\noindent
In the package's conventions $\Theta_3,\Thetab_3$ and $U$ correspond respectively to \texttt{\footnotesize \texttheta3, \texttheta b3} and \texttt{\footnotesize x3}. From the definition in \eqref{DQQbdef} and \eqref{opFirst} we have\footnote{In the package the derivatives \texttt{\scriptsize ChiralD} represent the derivatives $D$. When acting on the $t$ use \texttt{\scriptsize Chiral$\mathcal{D}$} instead ($\CD = $\texttt{\scriptsize \textcolor{black!60}{[esc]}scD\textcolor{black!60}{[esc]}}). Also note that, when acting on the $t$, the operators $D$ are sent to $\overbar{\CD}$ and $\Db$ to $\CD$. See \eqref{CDdefinition}.}
\vspace{-2ex}
\codein{
Dppt = \parbox{1.5cm}{\[\frac{\mbox{app}}{\mbox{(j+1)}^2}\]} Chiral$\CD$bp[Chiral$\CD$p[tO\straightphi\straightphi b, \texteta 1, \texttheta3, 
     x3], \texteta b1, \texttheta3, x3] 
}
\vspace{-8ex}
\codeno{\qquad    
\parbox{1.5cm}{\[+\,\frac{\mbox{bpp}}{\mbox{(j+1)}^2}\]} Chiral$\CD$p[Chiral$\CD$bp[tO\straightphi\straightphi b, \texteta b1, \texttheta3, x3], \texteta1, \texttheta3, x3];
}

\vspace{-.5em}
\par\noindent
This is going to be a large expression, but we only need the lowest order, we can therefore set the $\theta$ variables to zero\footnote{In more complicated applications it is better to separate the various orders in $\theta_3$ and $\thetab_3$ and act on each piece only with the operators inside \texttt{\scriptsize Chiral$\CD$} (like $\partial_\eta\partial_\theta$, $\partial_\eta\partial_\rmx\thetab$ etc$\ldots$) that do not give zero when the $\theta$'s are suppressed. In the documentation of the package there is a worked out example.}
\codein{
Dppt =  SetToZero[Dppt, \{\texttheta3, \texttheta b3\}] // Reduction;
}
\vspace{-1.5em}
\par\noindent
Now we need to remember that $\phi$ is chiral and it satisfies a shortening condition. The shortening operators at the first two points pass through the $\CK_{\CO_1\CO_2}$ prefactor\footnote{See \eqref{shortening}.} and thus we need to impose\footnote{This instruction will generate an error but it can be ignored. It can be avoided by replacing the \texttt{\scriptsize $\mathcal{C}$[i]}'s with variables consisting in a single \texttt{\scriptsize Symbol}. It comes up because there is an automatic routine that sets \texttt{\scriptsize ConstQ} to true for the variables in \texttt{\scriptsize variables}, which works only if they are \texttt{\scriptsize Symbols}. This is not an issue because \texttt{\scriptsize ConstQ[$\CC$[i]]} is true by default after \texttt{\scriptsize SUSY3pf} is called.}
\codein{
chi = Compare[Chiral$\CD$p[tO\straightphi\straightphi b, \texteta2, \texttheta3, x3], 
 variables $\to$ Array[$\CC$,4]]
}
\vspace{-1.5em}
\par\noindent
The function \texttt{\footnotesize Compare} equates an expression to zero and solves for the variables given by taking into account all possible identities among tensor structures.\footnote{The problem of imposing the identities among tensor structures (Schouten identities) is actually bypassed by replacing numerical values to the various quantities. More details are given in the package documentation.} In principle one needs to impose chirality at point $3$ as well, but in this case the constraint at the second point is strong enough to fix everything. In more complicated situations this might not happen and then one has to use the results of \appendixname~\ref{app:thirdpoint}. Now we are ready to compare the derivative with the basis defined previously. In this simple case the basis has length one, but let us write the code in a way that easily generalizes for larger bases
\codein{
solQQbpp = Compare[ConstantsList[\!\![1\lsp;\lnsp;\llsp1]\!\!].tQQbOpp - (Dppt/.chi[\!\![1]\!\!])]
}
\vspace{-2.5em}
\codeout{
$\Big\{\!\!\Big\{$\!\!\,\,$\CA \to$ \parbox{2.8cm}{\[
\mbox{i\!\!\lnsp\i\lsp}\frac{\mbox{(2qO + j) }\mathcal{C}_{\mbox{\scriptsize1}}}{\mbox{(j+1)}^{\mbox{\scriptsize2}}}
\]
}$\Big\}\!\!\Big\}$
}
\vspace{-1.5em}
\par\noindent
where we defined $\CA = \lambda_{(Q\Qb\CO)\phi\phib}$. Here \texttt{\footnotesize ConstantsList} is just a list of standard constants. If no options are given to \texttt{\footnotesize Compare}, it will solve for those. The norm of this operator is known from \cite{Li:2014gpa}
\eqn{
\langle (Q\Qb\CO)^{++} (Q\Qb\CO)^{++}\rangle \propto -4 \,\frac{(2q_\CO + j)(2 q_\CO + j + 1)}{(j+1)^4}\,,
}[]
where the proportionality factor is the usual tensor structure for two-point functions. For the user's convenience it is saved in the package and it can be obtained by using the function \texttt{\footnotesize operatorNorm}. The computation of $(Q\Qb\CO)^{--}$ is almost identical and the computation of $(Q^2\Qb{}^2\CO)$ is structurally the same, but it is more involved to define the differential operator. They are all worked out in the notebook attached to this submission. In the end we get the expected result
\eqna{
\CG_{2q_\CO,j} = \;\lsp&g_{2q_\CO,j} + \frac{(2q_\CO+j)}{4(2q_\CO+j+1)}g_{2q_\CO+1,j+1} + \frac{(2q_\CO-j-2)}{4(2q_\CO-j-1)}g_{2q_\CO+1,j-1} \\& +\frac{(2q_\CO+j)(2q_\CO-j-2)}{16(2q_\CO+j+1)(2q_\CO-j-1)} g_{2q_\CO+2,j} \,.
}[]
The different factors of $2$ in the first two terms are due to the $(-2)^{-j}$ in the normalization of the blocks.

\section{Conclusions}

We defined superconformally covariant differential operators in $\CN=1,2$ superspace. They can be used to expand in components any superconformal primary that appears in a correlation function in superspace. The results for $\CN=2$ do not yet cover all possible cases so further work will be needed. In addition we generalized the results of~\cite{Buican:2014qla} by showing that the exotic chiral primaries are absent from any local SCFT. This conclusion followed from the fact that the three-point function of an exotic primary, its conjugate and the stress tensor $\CJ$ is incompatible with the Ward identities of $\CJ$.
\par
The action of the differential operators on three-point functions has been addressed in detail in order to render more systematic and efficient the computation of superdescendant OPE coefficients in terms of their superconformal primary; which was the main motivation behind this work. This opens the way to many interesting applications. The most important one is to compute superconformal blocks for spinning superprimaries, both protected and long. A place to start from could be the case of spin $1/2$~\cite{Karateev:2019pvw} where the conformal blocks are already available and the extra work needed to include supersymmetry is minimal. The aim of the Mathematica package that we introduced is to render all these tasks relatively straightforward and systematic.
\par
The approach we followed for $\CN=2$ supersymmetry was to expand a multiplet into $\CN=1$ multiplets of a chosen subalgebra. This is what is usually needed in most applications. However, we would like to point our focus to another possible direction. Recently it was discovered that $\CN \geq 2$ SCFTs possess an infinite dimensional symmetry characterized by a $2\llsp d$ chiral algebra~\cite{Beem:2013sza, Beem:2014rza, Lemos:2014lua}. This correspondence applies to correlators of a class of operators called ``Schur operators.'' Except for the case of chiral primaries ($B_1\bar{B}_1$) all Schur operators are conformal but not superconformal primaries. They have $\mathfrak{su}(2)_R$ eigenvalue $R/2$ with charges
\eqn{
2q = j+R\,,\qquad 2\qb = \jb + R\,,
}[]
$(j,\jb)$ being the spin. It could be interesting to extend this formalism including a way to relate a correlator of Schur operators to the correlator of its corresponding primary.

\ack{
I am supported by the Swiss National Science Foundation under grant no.\ PP00P2-163670 and by the Doc.Mobility program of the Swiss National Science Foundation. I would like to thank Perimeter Institute and the Simons Center for Geometry and Physics for their hospitality during the preparation of parts of this manuscript. I would also like to thank Andreas Stergiou and David Poland for discussions and Alessandro Vichi for valuable comments on the draft and discussions.
}

\appendices

\section{Details on notation and conventions}\label{app:notation}

We use lowercase greek letters $\alpha,\alphad$ for spinor indices, lowercase latin letters $i,j$ for operator labels and uppercase latin letters $I,J$ for $\mathfrak{su}(2)_R$ indices (only for $\CN=2$).
The bosonic supersymmetric interval is defined as\footnote{Here and in the following equations the expressions for $\CN=1$ are obtained by dropping the $I$ indices.}
\eqn{
(\rmx_{i\jb})_{\alpha\alphad} = (\rmx_{ij})_{\alpha\alphad} - 2i \lsp\theta_{I i\alpha}\thetab^I_{i\alphad} - 2i \lsp\theta_{I j\alpha}\thetab^I_{j\alphad} + 4i\lsp \theta_{I i\alpha}\thetab^I_{j\alphad}\,,
}[xijdef]
with $x_{ij}$ being a shorthand for $x_i-x_j$ and $\rmx_{\alpha\alphad}$ denoting the matrix $\sigma^\mu_{\alpha\alphad}\lsp x_\mu$. We can also define the matrix with upper indices and the Lorentz square as follows
\eqn{
(\tilde\rmx_{\jb i})^{\alphad\alpha} = - \epsilon^{\alpha\beta}\epsilon^{\alphad\betad} (\rmx_{i\jb})_{\beta\betad} \,,\qquad 
{x_{\jb i}}^2 = \tfrac12 (\rmx_{i\jb})_{\alpha\alphad}\,(\tilde\rmx_{\jb i})^{\alphad\alpha}\,.
}[]
The fermionic supersymmetric intervals are defined as
\eqn{
\theta_{I\,ij}^\alpha = \theta^\alpha_{Ii} - \theta^\alpha_{Ij}\,,\qquad
\thetab_{\alphad\,ij}^I = \thetab^I_{\alphad i} - \thetab^I_{\alphad j}\,.
}[thetaidef]
The chiral derivatives in superspace are defined in the standard way
\eqn{
D_{i\lsp\alpha}^I = \frac{\partial}{\partial \theta_{I\lsp i}^\alpha} + i  \sigma^\mu_{\alpha\alphad}\lsp\thetab_i^{I\lsp \alphad} \frac{\partial}{\partial x_i^\mu}\,,\qquad
\Db_{I\lsp i\lsp\alphad} = -\frac{\partial}{\partial \thetab_i^{I\lsp \alphad}} - i  \theta_{I\lsp i}^\alpha\sigma^\mu_{\alpha\alphad}\frac{\partial}{\partial x_i^\mu}\,.
}[chiralDDbdef]
The derivative $D_i$ yields zero when acting on $\rmx_{k\jb},\,k\neq i$ and similarly for $\Db_i$ when acting on $\rmx_{j\kb},\,k\neq i$. For three-point functions in $\CN=1,2$ superspace we encountered the superconformally covariant variables $X_3,\Theta_3^I$ and $\Thetab_{3\lsp I}$. Here is their definition
\eqna{
&\mathrm{X}_3 = \frac{\rmx_{3\bar{1}}\tilde{\rmx}_{\bar{1}2}\rmx_{2\bar{3}}}{{x_{\bar{1}3}}{\!}^2{x_{\bar{3}2}}{\!}^2}\,,\quad
\overbarUp{\mathrm{X}}_3 = - \frac{\rmx_{3\bar{2}}\tilde{\rmx}_{\bar{2}1}\rmx_{1\bar{3}}}{{x_{\bar{3}1}}{\!}^2{x_{\bar{2}3}}{\!}^2} = \mathrm{X}_3 ^\dagger\,,\quad \\
&
\Theta_3^I = i \left(\frac{\rmx_{3\bar{1}}\thetab_{31}^I}{{x_{\bar{1}3}}{\!}^2}-\frac{\rmx_{3\bar{2}}\thetab_{32}^I}{{x_{\bar{2}3}}{\!}^2}\right)\,,\quad \Thetab_{3\lsp I} = i \left(\frac{\theta_{31\lsp I}\,\rmx_{1\bar{3}}}{{x_{\bar{3}1}}{\!}^2}-\frac{\theta_{32\lsp I}\,\rmx_{2\bar{3}}}{{x_{\bar{3}2}}{\!}^2}\right)=(\Theta_3^I)^\dagger \,.
}[XThThbdef]
Similar objects $\mathrm{X}_i,\,\Theta_i^I,\,\Thetab_{i\lsp I}$, $i=1,2$, can be defined by a cyclic permutation of the points. We will further define
\eqn{
\mathrm{U}_3 = \lifrac12(\mathrm{X}_3+\overbarUp{\mathrm{X}}_3)\,.
}[Udef]
Also, note that $\mathrm{X}_3-\overbarUp{\mathrm{X}}_3 = 4i\lsp
\Theta_3^I\Thetab_{3\lsp I}$.
We can define chiral derivatives in the space of the $X_i,\Theta_i,\Thetab_i$ variables
\begin{subequations}\label{CDdef}
\begin{align}
\CD_{\alpha\lsp I} &= \frac{\partial}{\partial \Theta^{\alpha\lsp I}} - 2 i  \sigma^\mu_{\alpha\alphad}\lsp\Thetab^\alphad_I \frac{\partial}{\partial X^\mu}\,,&\qquad
\overbar{\CD}_{\alphad}^I &= -\frac{\partial}{\partial \Thetab^\alphad_I}\,,\label{CDdefX}
\\
\CD_{\alpha\lsp I} &= \frac{\partial}{\partial \Theta^{\alpha\lsp I}}\,,&\qquad
\overbar{\CD}_{\alphad}^I &= -\frac{\partial}{\partial \Thetab^\alphad_I} + 2 i  \lsp\Theta^{\alpha\lsp I}\sigma^\mu_{\alpha\alphad} \frac{\partial}{\partial \Xb^\mu}\,,
\\
\mathcal{D}_{\alpha\lsp I} &= \frac{\partial}{\partial \Theta^{\alpha\lsp I}} -  i  \sigma^\mu_{\alpha\alphad}\lsp\Thetab^\alphad_I \frac{\partial}{\partial U^\mu}\,,&\qquad
\overbar{\CD}_{\alphad}^I &= -\frac{\partial}{\partial \Thetab^\alphad_I} + i  \lsp\Theta^{\alpha\lsp I}\sigma^\mu_{\alpha\alphad} \frac{\partial}{\partial U^\mu}\,.
\end{align}
\end{subequations}
All three representations are equivalent, provided one regards as independent the variables that appear therein. Furthermore we can give a list of dual derivatives $\CQ_\alpha$, $\overbar{\CQ}_\alphad$ satisfying $[\CD,\CQ\} = 0$.
\begin{subequations}\label{CQdef}
\begin{align}
\mathcal{Q}_{\alpha\lsp I} &= \frac{\partial}{\partial \Theta^{\alpha\lsp I}} + 2 i  \sigma^\mu_{\alpha\alphad}\lsp\Thetab^\alphad_I \frac{\partial}{\partial X^\mu}\,,&\qquad
\overbar{\mathcal{Q}}_{\alphad}^I &= -\frac{\partial}{\partial \Thetab^\alphad_I}\,,
\label{CQdefThetab}\\
\CQ_{\alpha\lsp I} &= \frac{\partial}{\partial \Theta^{\alpha\lsp I}}\,,&\qquad
\overbar{\mathcal{Q}}_{\alphad}^I &= -\frac{\partial}{\partial \Thetab^\alphad_I} - 2 i  \lsp\Theta^{\alpha\lsp I}\sigma^\mu_{\alpha\alphad} \frac{\partial}{\partial \Xb^\mu}\,,
\label{CQdefTheta}\\
\CQ_{\alpha\lsp I} &= \frac{\partial}{\partial \Theta^{\alpha\lsp I}} +  i  \sigma^\mu_{\alpha\alphad}\lsp\Thetab^\alphad_I \frac{\partial}{\partial U^\mu}\,,&\qquad
\overbar{\CQ}_{\alphad}^I &= -\frac{\partial}{\partial \Thetab^\alphad_I} - i  \lsp\Theta^{\alpha\lsp I}\sigma^\mu_{\alpha\alphad} \frac{\partial}{\partial U^\mu}\,.
\end{align}
\end{subequations}
These operators are obtained from the former by letting $(X,\Theta,\Thetab) \leftrightarrow (-\Xb,-\Theta,-\Thetab)$, which is also the result of swapping $1\leftrightarrow2$ for $(X_3,\Theta_3,\Thetab_3)$, and by introducing an overall minus sign by convention. It is possible to derive some very useful identities that arise from acting with $D$ or $\Db$ on a function of $Z_3$. For $\CN=1$ we have~\cite{Osborn:1998qu}
\eqna{
D_{1\lsp\alpha} \,f(Z_3) &= -i \frac{(\rmx_{1\bar3})_{\alpha\alphad}}{{x_{\bar13}}^2}\,\epsilon^{\alphad\betad}\lsp\overbar{\CD}_\betad\,f(Z_3)\,,\\
\Db_{1\lsp\alphad} \,f(Z_3) &= -i \frac{(\rmx_{3\bar1})_{\alpha\alphad}}{{x_{\bar31}}^2}\,\epsilon^{\alpha\beta}\lsp\CD_\beta\,f(Z_3)\,.
}[DonF]
Similarly, for the second point we have
\eqna{
D_{2\lsp\alpha} \,f(Z_3) &= i \frac{(\rmx_{2\bar3})_{\alpha\alphad}}{{x_{\bar23}}^2}\,\epsilon^{\alphad\betad}\lsp\overbar{\CQ}_\betad\,f(Z_3)\,,\\
\Db_{2\lsp\alphad} \,f(Z_3) &= i \frac{(\rmx_{3\bar2})_{\alpha\alphad}}{{x_{\bar32}}^2}\,\epsilon^{\alpha\beta}\lsp\CQ_\beta\,f(Z_3)\,.
}[altDonF]
And for $\CN=2$ we have~\cite{Kuzenko:1999pi}
\eqna{
D_{1\lsp\alpha}^I \,f(Z_3) &= -i \frac{(\rmx_{1\bar3})_{\alpha\alphad}}{({x_{\bar13}}^2{x_{\bar31}}^2)^{1/2}}\,\hat{u}_J^{\phantom{J}\lsp I}(z_{31})\,\epsilon^{\alphad\betad}\lsp{\overbar{\CD}_\betad}^J\,f(Z_3)\,,\\
\Db_{1\lsp\alphad\lsp I} \,f(Z_3) &= -i \frac{(\rmx_{3\bar1})_{\alpha\alphad}}{({x_{\bar13}}^2{x_{\bar31}}^2)^{1/2}}\,\hat{u}_I^{\phantom{I}\lsp J}(z_{13})\,\epsilon^{\alpha\beta}\lsp\CD_{\beta\lsp J}\,f(Z_3)\,,
}[DonFN2]
having defined $u_I^{\phantom{I} J}$ in \eqref{uijdef}. Similarly, for the second point we have
\eqna{
D_{2\lsp\alpha}^I \,f(Z_3) &= i \frac{(\rmx_{2\bar3})_{\alpha\alphad}}{({x_{\bar23}}^2{x_{\bar32}}^2)^{1/2}}\,\hat{u}_J^{\phantom{J}\lsp I}(z_{32})\,\epsilon^{\alphad\betad}\lsp{\overbar{\CQ}_\betad}^J\,f(Z_3)\,,\\
\Db_{2\lsp\alphad\lsp I} \,f(Z_3) &= i \frac{(\rmx_{3\bar2})_{\alpha\alphad}}{({x_{\bar23}}^2{x_{\bar32}}^2)^{1/2}}\,\hat{u}_I^{\phantom{I}\lsp J}(z_{23})\,\epsilon^{\alpha\beta}\lsp\CQ_{\beta\,J}\,f(Z_3)\,.
}[altDonFN2]

\section{Acting on different points}\label{app:thirdpoint}

The formulas shown in this paper only consider the application of the differential operators at the first point. Due to the symmetry of exchanging the first two points, it is very easy to derive similar formulas for the second one. It suffices to use \eqref{altDonF} instead of \eqref{DonF}. The result amounts to simply replacing all $\CD$'s to $\CQ$'s and $\overbar{\CD}$'s to $\overbar{\CQ}$'s and including an extra minus sign for every derivative. The result concerning the commutativity with the prefactor clearly holds as well thanks to the symmetry of $\CK_{\CO_1\CO_2}$ upon exchanging the first two operators.
\par
On the other hand, the point $\bfz_3$ is treated differently by the parametrization of \eqref{GeneralThreepf}. As a consequence it is not possible to apply a differential operator on $\CO_3$ using the formulas shown here. Fortunately there is a way to switch between different parametrizations by working on the $t$ only. For simplicity, we will assume all three operators to be $\mathfrak{su}(2)$ singlets. We have
\eqn{
\langle \CO_1(\bfz_1)\CO_2(\bfz_2)\CO_3(\bfz_3)\rangle = \CK_{\CO_1\CO_2}(\bfz_{1,2},z_3) \,t_{\CO_3}^{\CO_1\CO_2}(Z_3)\,,
}[]
where some arguments of the various objects are not shown for brevity. Recall $\bfz = z,\eta,\etab$, $z = x,\theta,\thetab$. By cyclically permuting the operators (which is a trivial operation) we get
\eqn{
\langle \CO_2(\bfz_2)\CO_3(\bfz_3) \CO_1(\bfz_1)\rangle = \CK_{\CO_2\CO_3}(\bfz_{2,3},z_1) \,t_{\CO_1}^{\CO_2\CO_3}(Z_1)\,,
}[]
having used $Z_3|_{\{1,2,3\} \to \{2,3,1\}} = Z_1$. The reduced three-point functions $t$ satisfy a scaling property
\eqn{
t_{\CO_3}^{\CO_1\CO_2}(\lambda\lambdab X,\lambda\Theta,\lambdab\Thetab) = \lambda^{2a}\lambdab^{2\ab} \lsp t_{\CO_3}^{\CO_1\CO_2}(X,\Theta,\Thetab)\,,
}[scalingt]
with
\eqna{
a &= \tfrac13 (q_3 - \qb_1 - \qb_2) + \tfrac23(\qb_3-q_1-q_2)\,,\\
\ab &= \tfrac23 (q_3 - \qb_1 - \qb_2) + \tfrac13(\qb_3-q_1-q_2)\,.
}[]
Let us be now more precise with the arguments of the $t$. We will denote it as
\eqn{
t^{\CO_i\CO_j}_{\CO_k}(Z;\eta_i,\etab_i;\eta_j,\etab_j;\eta_k,\etab_k)\,,
}[]
implying that the spinors are associated to the operator with the matching label. Following \cite{Osborn:1998qu} we can then write the following formula that relates the two
\eqn{
t_{\CO_1}^{\CO_2\CO_3}(Z;\eta_2,\etab_2;\etab_3\overbarUp{\mathrm{X}},\mathrm{X}\eta_3;\eta_1,\etab_1) = C(X,\Xb)\, t_{\CO_3}^{\CO_1\CO_2}(Z; \etab_1\mathrm{X},\overbarUp{\mathrm{X}}\eta_1;\eta_2,\etab_2;\eta_3,\etab_3)\,.
}[onetothree]
with
\eqn{
C(X,\Xb) =  \frac{(-1)^{\llsp j_1 + j_2+\jb_2+\jb_3}
}{{X}\rule{0pt}{9.70pt}^{2(a+\qb_2)+\jb_1-j_3}\lsp \Xb^{2(\ab+q_2)+j_1-\jb_3}}\,.
}[]
\par
The matching involves only quantities in the $Z$ space. If one wants to have the $\mathrm{X}\eta$ replacements only in one side,  it is possible to consider equation \onetothree with, for instance, $\eta_1 \to \overbarUp{\mathrm{X}}\etab_1$ and $\etab_1\to\eta_1\mathrm{X}$ and then use simply
\eqn{
\overbarUp{\mathrm{X}}\overbarUp{\mathrm{X}} \etab_1 = - \Xb^2 \etab_1\,,\qquad\eta_1\mathrm{X}\mathrm{X}= - X^2 \eta_1\,.
}[]
The factors of $X^2$ and $\Xb{}^2$ can be then taken out using the scaling in $j_1,\jb_1$. Often it is more convenient to express the correlator in terms of $U$ rather than $X$ and $\Xb$. The relation is very simple
\eqn{
\mathrm{X}_{\alpha\alphad}= \mathrm{U}_{\alpha\alphad} + 2 i \lsp\Theta_\alpha\Thetab_\alphad\,,\qquad
\overbarUp{\mathrm{X}}_{\alpha\alphad}= \mathrm{U}_{\alpha\alphad} - 2 i \lsp\Theta_\alpha\Thetab_\alphad\,.
}[]
A function that solves \eqref{onetothree} has been implemented in the Mathematica package of Sec.~\ref{sec:mathematica} under the name of \texttt{\footnotesize permuteCyclic}.

\section{Superspace expansion}\label{app:details}

The expansion in $\CN=1$ superspace of a general superfield reads

\eqn{
\CO(\bfz) \big|_{\theta\,=\,\thetab\,=\,0} = O(\bfx)\,.
}[lowest]

\eqna{
\CO \big|_{\theta,\thetab} =
 &\phantom{\,-\,} i\lsp\theta\partial_\eta\,(QO)^+ + \tfrac{j}{j+1}\,i\lsp\theta\eta\,(QO)^- \\
&-i\lsp\thetab\partial_\etab\,(\Qb O)^+ + \tfrac{\jb}{\jb+1}\,i\lsp\thetab\etab\,(\Qb O)^-\,.
}[first]

\eqn{\CO\big|_{\theta^2,\thetab^2} = \tfrac14\lsp\theta^2(Q^2O)  + \tfrac14\lsp\thetab^2(\Qb{}^2O)\,. 
}[thetasq]

\eqna{\CO \big|_{\theta\thetab} =
&-\lsp\theta\partial_\eta\,\thetab\partial_\etab \big((Q\Qb O)^{++} -i c_1\,\eta\lsp\partial_\rmx\etab\,O\big)\\
&-\tfrac{j}{j+1}\,\theta\eta\,\thetab\partial_\etab \big((Q\Qb O)^{-+}-i c_2\,\partial_\eta\partial_\rmx\etab\,O\big)\\
&+\tfrac{\jb}{\jb+1}\,\theta\partial_\eta\,\thetab\etab \big((Q\Qb O)^{+-}-i c_3\,\eta\lsp\partial_\rmx\partial_{\etab}\,O\big)\\
&+\tfrac{j\jb}{(j+1)(\jb+1)}\,\theta\eta\,\thetab\etab \big((Q\Qb O)^{--}-i c_4\,\partial_{\eta}\partial_\rmx\partial_{\etab}\,O\big)\,.
}[thetathetab]

\eqna{\CO \big|_{\theta^2\thetab, \thetab^2\theta} = 
&-\tfrac{i}4 \theta^2\,\thetab\partial_\etab\,\big((Q^2\Qb O)^+ -i c_5\,\partial_\eta\partial_\rmx\etab \,(QO)^+ -i c_6\,\eta\lsp\partial_\rmx\etab\,(QO)^-\big)\\
&+\tfrac{i}4 \tfrac{\jb}{\jb+1} \thetab^2\,\thetab\etab\,\big((Q^2\Qb O)^- -i c_7\,\partial_\eta\partial_\rmx\partial_\etab \,(QO)^+ -i c_8\,\eta\lsp\partial_\rmx\partial_{\etab}\,(QO)^-\big)\\
&+\tfrac{i}4 \thetab^2\,\theta\partial_\eta\,\big((\Qb{}^2 Q O)^+ -i \bar{c}_5\,\eta\lsp\partial_\rmx\partial_{\etab} \,(\Qb O)^+ -i \bar{c}_6\,\eta\lsp\partial_\rmx\etab\,(\Qb O)^-\big)\\
&+\tfrac{i}4 \tfrac{j}{j+1} \thetab^2\,\theta\eta\,\big((\Qb{}^2 Q O)^- -i \bar{c}_7\,\partial_\eta\partial_\rmx\partial_\etab \,(\Qb O)^+ -i \bar{c}_8\,\partial_{\eta}\partial_\rmx\etab\,(\Qb O)^-\big)\,.
}[thetasqthetab]

\eqna{\CO \big|_{\theta^2\thetab^2} =
\tfrac1{16}\,\theta^2\lsp\thetab^2\,\big(&
(Q^2\Qb{}^2O) \\
&- i c_9\,\partial_\eta\partial_\rmx\partial_\etab\, (Q\Qb O)^{++} - i c_{10}\,\eta\lsp\partial_\rmx\partial_{\etab} \,(Q\Qb O)^{-+}\\
&- i c_{11}\,\partial_\eta\partial_\rmx \etab\, (Q\Qb O)^{+-} - i c_{12}\,\eta\lsp\partial_\rmx \etab \,(Q\Qb O)^{--}\\
&-c_{13} \,\partial^2_x\,O - c_{14}\,\partial_\eta\partial_\rmx\partial_\etab\,\lsp \eta\lsp\partial_\rmx \etab\,O
\big)\,.
}[thetasqsq]
\noindent
The coefficients $c_1$ through $c_4$ agree with \cite{Li:2014gpa}. Instead $c_5$ through $c_{12}$ and $c_{14}$ differ by a simple normalization (cfr. \cite[(A.8--10)]{Li:2014gpa}). Finally $c_{13}$ is a bit different
\eqn{
c_{13}^{\mathrm{here}} = - 4\lsp (2+j+\jb)\big(c_{13}^{\mathrm{there}} + 8\lsp c_{14}^{\mathrm{there}}\big)\,.
}[]
The reason is because they define $\partial_x^2 O$ to be
\eqn{
- \tfrac12\, \partial_\eta\partial_\rmx\partial_\etab\,\lsp \eta\lsp\partial_\rmx \etab\,O + \tfrac12\, \eta\lsp\partial_\rmx\etab\,\lsp \partial_\eta\partial_\rmx \partial_\etab\,O = \tfrac12(2+j+\jb)\, \partial_\mu\partial^\mu O\,,
}[]
whereas we define it simply as $\partial_\mu\partial^\mu O$. The ancillary file attached to this submission contains the expressions of \eqref{abdef}, \eqref{cdedef} and \eqref{fidef} in terms of the coefficients $c_i$.

\section{Some identities for the superspace derivatives}\label{app:identities}

Let us denote in this way the following shifts of the quantum numbers in $\CK_{\CO_1\CO_2}$
\eqn{
\CK_{s_\pm\CO_1\CO_2} = \CK_{\CO_1\CO_2}\raisebox{-1.3ex}{\Big |}{}_{\substack{q_1\to q_1+1/2\\j_1\to j_1\pm1\hspace{1.6ex}}}
\,,\qquad
\CK_{\bar{s}_\pm\CO_1\CO_2} = \CK_{\CO_1\CO_2}\raisebox{-1.3ex}{\Big |}{}_{\substack{\qb_1\to \qb_1+1/2\\\jb_1\to \jb_1\pm1\hspace{1.6ex}}}\,.
}[shiftss]
Note that this does not correspond to the shifts that follow from applying $Q$ or $\Qb$ on $\CO_1$. Those have been defined in \eqref{shiftsQ}. This definition simply happens to be convenient for the formulas that will follow. The factors of $j$ are inserted to cancel the contributions from the factorials at the denominator of \eqref{Kdef}.
\par
The first set of formulae that we need is the action of the first order derivatives on $\CK_{\CO_1\CO_2}$. The quantum numbers of $\CO_1$ will be denoted as $q,\qb,j$ and $\jb$. Since in some cases $\CO_1$ might be a superdescendant we will denote as $j_0$ and $\jb_0$ the spin labels of the superconformal primary. They will show up because they are inside the definition of the differential operators (see \eqref{opFirst}).
\fourseqn{
\DQm\,\CK_{\CO_1\CO_2} &= -2(2q - j -2)\frac{1}{j_0}\,\CK_{s_-\CO_1\CO_2}\,\thetab_1\partial_\chib\,,
}[]{
\DQp\,\CK_{\CO_1\CO_2} &= -2\frac{2q+j}{j_0+1}\,\CK_{s_+\CO_1\CO_2}\,\chib\thetab_1\,,
}[]{
\DQbm\,\CK_{\CO_1\CO_2} &= -2(2\qb - \jb -2)\frac{1}{\jb_0}\,\CK_{\bar{s}_-\CO_1\CO_2}\,\partial_\chi\theta_1\,.
}[]{
\DQbp\,\CK_{\CO_1\CO_2} &= -2\frac{2\qb+\jb}{\jb_0+1}\,\CK_{\bar{s}_+\CO_1\CO_2}\,\theta_1\chi\,.
}[][DonK]
Next we need the action on $\CK_{\CO_1\CO_2}$ times a Grassmann variable. Below, the $\chi$ and $\chib$ that appear on both sides of the equalities can also be replaced by $\partial_\chi$ or $\partial_\chib$ being careful with the signs.\footnote{In our conventions the replacement is $\chi^\alpha \to \partial_{\chi_{\alpha}}$ and $\chi_\alpha \to - \partial_{\chi^\alpha}$. The same holds for $\chib$.}
\fourseqn{
\DQm\,\CK_{\CO_1\CO_2} \, \theta_1\chi &= (\DQm\,\CK_{\CO_1\CO_2}) \, \theta_1\chi -  \frac{i}{j_0}\,\CK_{s_-\CO_1\CO_2}\,\chi\lsp\rmx_{1\bar3}\partial_{\chib}\,,
}[]{
\DQp\,\CK_{\CO_1\CO_2} \,\theta_1\chi &= (\DQp\,\CK_{\CO_1\CO_2}) \,\theta_1\chi + \frac{i}{j_0+1}\,\CK_{s_+\CO_1\CO_2}\,\chi\lsp\rmx_{1\bar3}\chib\,,
}[]{
\DQbm\,\CK_{\CO_1\CO_2} \,\chib\thetab_1 &= (\DQbm\,\CK_{\CO_1\CO_2}) \, \chib\thetab_1 +  \frac{i}{\jb_0}\,\CK_{\bar{s}_-\CO_1\CO_2}\,\partial_\chi\rmx_{3\bar1}\chib\,,
}[]{
\DQbp\,\CK_{\CO_1\CO_2} \,\chib\thetab_1 &=   (\DQbp\,\CK_{\CO_1\CO_2}) \,\chib\theta_1 - \frac{i}{\jb_0+1}\,\CK_{\bar{s}_+\CO_1\CO_2}\,\chi\lsp\rmx_{3\bar1}\chib\,.
}[][DonKtheta]
Finally we will also need the action of the derivatives on the $t$. These equations will make use of a different definition for the shifts, which can be found in \eqref{shiftsQ}. The result is
\twoseqn{
\CK_{\CO_1\CO_2}\,\DQ\pm\,t^{\CO_1\CO_2}_{\CO_3} &=  \CK_{(Q\CO_1)^\pm\CO_2}\,\CDQ\pm\,t^{\CO_1\CO_2}_{\CO_3} \,,
}[]{
\CK_{\CO_1\CO_2}\,\DQb\pm\,t^{\CO_1\CO_2}_{\CO_3} &= \CK_{(\Qb\CO_1)^\pm\CO_2}\,\CDQb\pm\,t^{\CO_1\CO_2}_{\CO_3} \,,
}[][KDont]
where we have used the derivatives defined in \eqref{CopFirst}.
\par
All these results can be easily proven by first applying the differential operators on simple terms such as $\rmx_{i\jb}$ and ${x_{\ib j}}^2$ and then working our way up to more  complicated expressions. The main trick used involves shifting the labels of the prefactor. When the labels are shifted down that means that a derivative acted on $\eta$ or $\etab$, bringing down a $\partial_\chib$ or $\partial_\chib$ respectively. When the labels are shifted up that means that we have introduced an extra auxiliary spinor $\chi$ or $\chib$ on which the added derivative can act and reproduce the needed expression.

\renewcommand{\ttdefault}{cmtt}
\newpage

\Bibliography

\end{document}